\documentstyle[12pt,psfig]{ioplppt} 

\def\bbbq{{\mathchoice {\setbox0=\hbox{$\displaystyle\rm Q$}\hbox{\raise
0.15\ht0\hbox to0pt{\kern0.4\wd0\vrule height0.8\ht0\hss}\box0}}
{\setbox0=\hbox{$\textstyle\rm Q$}\hbox{\raise
0.15\ht0\hbox to0pt{\kern0.4\wd0\vrule height0.8\ht0\hss}\box0}}
{\setbox0=\hbox{$\scriptstyle\rm Q$}\hbox{\raise
0.15\ht0\hbox to0pt{\kern0.4\wd0\vrule height0.7\ht0\hss}\box0}}
{\setbox0=\hbox{$\scriptscriptstyle\rm Q$}\hbox{\raise
0.15\ht0\hbox to0pt{\kern0.4\wd0\vrule height0.7\ht0\hss}\box0}}}}
\def\bbbz{{\mathchoice {\hbox{$\sf\textstyle Z\kern-0.4em Z$}}
{\hbox{$\sf\textstyle Z\kern-0.4em Z$}}
{\hbox{$\sf\scriptstyle Z\kern-0.3em Z$}}
{\hbox{$\sf\scriptscriptstyle Z\kern-0.2em Z$}}}}

\begin{document}

\jl{1}

\title{On the accuracy of the semiclassical trace formula}

\author{Harel Primack
          \footnote{Present address: Fakult\"{a}t f\"{u}r Physik,
                    Albert-Ludwids Universit\"{a}t,
                    Hermann-Herder-Str.\ 3,
                    D-79104 Freiburg, Germany}
        and Uzy Smilansky}

\address{{Department of  Physics of Complex Systems,} \\
         {The Weizmann Institute of Science,
          Rehovot 76100, Israel}}

\date{\today}

\begin{abstract}
%
The semiclassical trace formula provides the basic construction from
which one derives the semiclassical approximation for the spectrum of
quantum systems which are chaotic in the classical limit. When the
dimensionality of the system increases, the mean level spacing
decreases as $\hbar^d$, while the semiclassical approximation is
commonly believed to provide an accuracy of order $\hbar^2$, {\it
independently} of $d$. If this were true, the semiclassical trace
formula would be limited to systems in $d \leq 2$ only. In the present
work we set about to define proper measures of the semiclassical
spectral accuracy, and to propose theoretical and numerical evidence
to the effect that the semiclassical accuracy, measured in units of
the mean level spacing, depends only weakly (if at all) on the
dimensionality.  Detailed and thorough numerical tests were performed
for the Sinai billiard in $2$ and $3$ dimensions, substantiating the
theoretical arguments.
\end{abstract}


\maketitle

\section{Introduction}
\label{sec:intro}
%
The semiclassical analysis has proven to be a very useful tool in the
field of ``quantum chaos'' as well as in many other fields.
Nevertheless, one should bear in mind, that it only {\em approximates}
the true quantal solution. Thus, it is imperative to know what are the
{\em errors} which are inherent to the semiclassical approximation,
and whether they could be considered as sufficiently small for the
problem at hand.

We shall focus our attention on one particular application of the
semiclassical approximation: The calculation of the energy spectra
of classically chaotic systems. The analytical tool that is used for
this purpose is the semiclassical Gutzwiller trace formula
\cite{Gut90} which expresses the {\it quantum} spectral density in
terms of {\it classical} quantities, and in particular the actions and
stabilities of classical periodic orbits. The trace formula was used,
among other things, to explain and discuss spectral statistics and
their relation to the universal predictions of Random Matrix Theory
(RMT) \cite{Ber85,BK96}. However, a prerequisite for the use of the
semiclassical approximation to compute short--range statistics is that
it is able to reproduce the exact spectrum within an error which is
comparable to or less than the mean level spacing!  This is a
demanding requirement, and quite often, the ability of the
semiclassical approximation to reproduce precise levels for
high--dimensional systems is doubted, and on the following
grounds. The mean level spacing depends on the dimensionality (number
of freedoms) of the system, and it is ${\cal O}(\hbar^d)
\cite{LLqm}$. Gutzwiller \cite{Gut90} uses an argument by Pauli
\cite{Pau51} to show that in general the error margin for the
semiclassical approximation scales as ${\cal O}(\hbar^2)$ {\em
independently of the dimensionality}.  Applied to the trace formula,
the expected error in units of the mean spacing, which is the figure
of merit in the present context, is therefore expected to be ${\cal
O}(\hbar^{2-d})$. We shall refer to this as the ``traditional
estimate". It sets $d=2$ as a critical dimension for the applicability
of the semiclassical trace formula and hence for the validity of the
conclusions which are drawn from it. The few systems in $d>2$
dimensions which were numerically investigated display spectral
statistics which adhere to the predictions of RMT as accurately as
their counterparts in $d=2$ \cite{PS95,Pri97,Pro97}. Thus, the
``traditional estimate'' cannot be entirely correct in the present
context, and we shall explain the reasons why it is inadequate when we
discuss other error bounds in the next section.

It is rather surprising that the problem of the accuracy of the
semiclassical trace formula was rather rarely touched upon in the
literature. Gutzwiller quotes the ``traditional'' estimate of ${\cal
O}(\hbar^{2-d})$ which was discussed above \cite{Gut90,Gut89}. Gaspard
and Alonso \cite{GA93} and Alonso and Gaspard \cite{AG93} derived
explicit (generic) $\hbar$ corrections for the periodic orbit terms in
the trace formula, but did not investigate their effect on the
semiclassical accuracy of energy levels. Diffraction corrections were
discussed in the context of the trace formula by Vattay, Wirzba and
Rosenqvist \cite{VWR94} and by Primack et al. \cite{PSSU97}. Also in
these works the focus is on the corrections to individual periodic
orbit terms rather than on the overall effect on energy levels.
Boasman \cite{Boa94} studied the accuracy of the Boundary Integral
Method (BIM) \cite{BW84} for 2D billiards in the case that the exact
kernel is replaced by its semiclassical asymptotic approximation. He
finds that the resulting error is of the same magnitude as the mean
spacing, which is consistent with the traditional estimate. However,
Boasman's work does not refer directly to the trace formula and to the
periodic orbits contributions. The works of Bleher \cite{Ble91,Ble94}
and of Prosen and Robnik \cite{PR93} discuss the accuracy of the
semiclassical approximation in the integrable case.

The purpose of this work is to address the subject of the accuracy of
the semiclassical trace formula conceptually, theoretically and
numerically. We shall be interested in particular in the dependence of
the semiclassical error on the {\em dimension} of the system. To do
so, we shall have to start by developing the basic concepts and define
the measures we use for a quantitative estimate of the spectral error
(section \ref{sec:accuracy-measures}). The accuracy of the
semiclassical approximation of the quantal energy spectrum will be
then studied via the dual classical spectrum of actions and
stabilities of periodic orbits (time spectrum). This will enable us to
use our data base of quantum levels and periodic orbits for the Sinai
billiards in two and three dimensions for a direct evaluation of the
semiclassical error (section \ref{sec:sinai-numerics}). We shall
summarize the paper and discuss a few relevant points in section
\ref{sec:discussion}.

\section{Measures of the semiclassical error}
\label{sec:accuracy-measures}
%
In order to define a proper error measure for the semiclassical
approximation of the energy spectrum one has to clarify a few
issues. In contrast with the EBK quantization which gives an {\em
explicit} formula for the spectrum, the semiclassical spectrum for
chaotic systems is {\em implicit} in the trace formula, or in the
semiclassical expression for the spectral determinant.  To extract the
semiclassical spectrum we recall that the exact spectrum, $\{E_n \}$,
can be obtained from the exact counting function:
\begin{equation}
  N(E) \equiv \sum_{n=1}^{\infty} \Theta ( E - E_n ) \ ,
  \label{eq:qmstair}
\end{equation}
by solving the equation
\begin{equation}
  N(E_n) = n - \frac{1}{2} \; ,
  \; \; n = 1, 2, \ldots \; \; \; .
  \label{eq:qc}
\end{equation}
In the last equation, an arbitrarily small amount of smoothing must be
applied to the Heavyside function. In complete analogy, one obtains
the semiclassical spectrum $\{E_n^{\rm sc}\}$ as \cite{AM96}:
\begin{equation}
  N_{\rm sc}(E_n^{\rm sc}) = n - \frac{1}{2} \; ,
  \; \; n = 1, 2, \ldots \; \; \; ,
  \label{eq:scqc}
\end{equation}
where $N_{\rm sc}$ is a semiclassical approximation of $N$. Note that
$N_{\rm sc}$ with which we start is not necessarily a sharp counting
function. However, once $\{E_n^{\rm sc}\}$ is known, we can
``rectify'' the smooth $N_{\rm sc}$ into the sharp counting function
$N^{\#}_{\rm sc}$ \cite{BK96}:
\begin{equation}
  N^{\#}_{\rm sc}(E) \equiv 
  \sum_{n=1}^{\infty} \Theta ( E - E_n^{\rm sc} ) \, .
  \label{eq:nsc-sharp}
\end{equation}
The simplest choice for $N_{\rm sc}$ is the Gutzwiller trace formula
\cite{Gut90} truncated at the Heisenberg time, which is what we shall 
use in the present paper. Alternatively, one can start from the
regularized Berry--Keating Zeta function $\zeta_{\rm sc}(E)$
\cite{Kea93}, and define $N_{\rm sc} = (1/\pi) \, \mbox{Im} \, \log \,
\zeta_{\rm sc} (E+i0)$, in which case $N_{\rm sc} = N^{\#}_{\rm sc}$.

Next, in order to define a quantitative measure of the semiclassical
error, one should establish a {\em correspondence} between the quantal
and the semiclassical levels, namely, one should identify the
semiclassical counterparts of the exact quantum levels. In classically
chaotic systems, for which the Gutzwiller trace formula is applicable,
the only constant of the motion is the energy. This is translated into
a single ``good'' quantum number in the quantum spectrum, which is the
ordinal number of the levels when ordered by their magnitude. Thus,
the only correspondence which can be established between the exact
spectrum $\{E_n\}$ and its semiclassical approximation, $\{E^{\rm
sc}_n\}$, is
\begin{equation}
  E_n \longleftrightarrow E^{\rm sc}_n \; .
\end{equation}
This is to be contrasted with integrable systems, where it is
appropriate to compare the exact and approximate levels which have the
same quantum numbers.

The scale on which the accuracy of semiclassical energy levels should
be measured depends, in general, on the problem at hand. The most
natural choice, however, is the mean level spacing $(\bar d(E))^{-1}$
where $\bar{d}$ is the smooth density of states. The semiclassical
error of the n'th level is therefore measured by \cite{Boa94}:
\begin{equation}
  \epsilon_{n} \equiv
  \bar{d}(E_n) \left( E_{n} - E^{\rm sc}_{n} \right) .
\end{equation}
A more useful and significant measure is the average of $\epsilon_{n}$
over an energy interval $\Delta E$ centered at $E$, which contains a
large number of quantum energies, but which must be so small that both
the classical dynamics and the mean density of states remain
approximately constant. This energy averaging will be denoted by
triangular brackets $\langle \cdot\rangle_E$ in the sequel. If the
semiclassical mean density $\bar{d}_{\rm sc}$ agrees with $\bar{d}$ to
a high precision, then obviously $\epsilon(E) \equiv \langle
\epsilon_{n} \rangle_{E}= 0$. In this case $\epsilon(E)$ is not a
useful error measure. For billiard systems this is always the case,
since the mean spectral density can be written as an asymptotic series
with explicitly known coefficients
\cite{BB70,BH76,BH94}. For general systems, only the leading
Weyl term is explicitly known. Two appropriate measures which are
sensitive to the accuracy of the fluctuating parts of the level
densities are the mean absolute error:
\begin{equation}
  \epsilon^{\rm (1)}(E)   \equiv
  \langle \  \bar{d}(E_n) \left| E_{n} - E^{\rm sc}_{n} \right| \
    \rangle_{E}
\end {equation}
and the variance:
\begin{equation}
  \epsilon^{\rm (2)}(E) \equiv
  \langle \ \left ( \bar{d}(E_n) 
    \left( E_{n} - E^{\rm sc}_{n} \right ) \right)^2 \
  \rangle_{E} \ .
\end{equation}

Having defined the spectral error measures, let us apply them and try
to get some estimates of the semiclassical error. In the introductory
section we have mentioned the ``traditional'' estimate of the
semiclassical error. Gutzwiller \cite{Gut90,Gut89} shows, based on
Pauli \cite{Pau51}, that the inherent error in the semiclassical
(Van-Vleck) approximation of the quantal time propagator scales like
${\cal O}(\hbar^2)$. Since the energy levels are the temporal Fourier
components of the propagator, it is plausible to assume that they have
the same degree of accuracy:
\begin{equation}
  E_{n} - E^{\rm sc}_{n} = {\cal O}(\hbar^2) \; \; \;
  \mbox{``traditional''}.
  \label{eq:common}
\end{equation}
(Strictly speaking, this is an upper bound.) The mean density of
energy levels, for a general $d$-dimensional system is asymptotically
given by Weyl's formula \cite{LLqm}:
\begin{equation}
  \bar{d}(E) \approx \frac{\omega(E)}{h^d} = {\cal O}(\hbar^{-d})
  \label{eq:dos}
\end{equation}
where $\omega(E)$ is the measure of the energy surface in the
classical phase space.
Hence,
\begin{equation}
  \epsilon^{\rm traditional} = {\cal O}(\hbar^{2-d})
  \longrightarrow
  \left\{
  \begin{array}{lcc}
     {\rm const} & , & d = 2 \\
     \infty      & , & d \ge 3
  \end{array}
  \right. \; \; \;
  \mbox{as } \ \  \hbar \rightarrow 0.
  \label{eq:epstrad}
\end{equation}
That is, the semiclassical approximation is (marginally) accurate in
2 dimensions, but it fails to predict accurate energy levels for 3
dimensions or more.

One may get a different estimate of the semiclassical error, if the
Gutzwiller Trace Formula (GTF) is used as a starting point. Suppose
that we have calculated $N_{\rm sc}$ to a certain degree of precision,
and we compute from it the semiclassical energies using
(\ref{eq:scqc}). The quality of this approximation can be estimated if
the leading corrections $\Delta N_{\rm sc}$ are also included and the
resulting energy differences $\delta_n$ are evaluated. We thus need to
consider:
\begin{equation}
  N_{\rm sc}(E_n^{\rm sc}+\delta_n) +
  \Delta N_{\rm sc}(E_n^{\rm sc}+\delta_n) =
  n - \frac{1}{2} \, .
  \label{eq:impscqc}
\end{equation}
Combining (\ref{eq:scqc}) and (\ref{eq:impscqc}) we get to first order
in $\delta_n$:
\begin{equation}
  \delta_n \approx \frac{\Delta N_{\rm sc}(E_n^{\rm sc})}
  {\partial N_{\rm sc}(E_n^{\rm sc}) / \partial E} \approx
  \frac{\Delta N_{\rm sc}(E_n^{\rm sc})}{\bar{d}(E_n^{\rm sc})}.
\end{equation}
In the above we assumed that the fluctuations of $N_{\rm sc}$ around
its average are not very large. Thus,
\begin{equation}
  \epsilon^{\rm GTF} \approx
  \bar{d}(E_n^{\rm sc}) \delta_n \approx
  \Delta N_{\rm sc}(E_n^{\rm sc}).
\end{equation}

Let us apply the above formula and consider the case in which we take
for $N_{\rm sc}$ its mean part $\bf\bar{\it N}$, and that we include
in $\bf\bar{\it N}$ terms of order up to (and including) $\hbar^{-m},
m \leq d$. For $\Delta N_{\rm sc}$ we use both the leading correction
to $\bf\bar{\it N}$ and the leading order periodic orbit sum which is
(formally) of order $\hbar^0$. Hence,
\begin{equation}
  \epsilon^{\rm GTF}_{\bar{N}} =
  {\cal O}(\hbar^{-m+1}) + {\cal O}(\hbar^0) =
  {\cal O}\left( \hbar^{\min(-m+1,\,0)} \right).
  \label{eq:tf-estimate}
\end{equation}
We conclude, that approximating the energies only by the mean counting
function $\bf\bar{\it N}$ up to (and not including) the constant term,
is already sufficient to obtain semiclassical energies which are
accurate to ${\cal O}(\hbar^0) = {\cal O}(1)$ with respect to the mean
density of states. Note again, that no periodic orbits were included
in $N_{\rm sc}$. Including less terms in $\bf\bar{\it N}$ will lead to
a diverging semiclassical error, while more terms will be masked by
the periodic orbit (oscillatory) term. One can do even better if one
includes in $N_{\rm sc}$ the smooth terms up to and including the
constant term (${\cal O}(\hbar^0)$) together with the leading order
periodic orbit sum which is formally also ${\cal O}(\hbar^0)$. The
semiclassical error is then:
\begin{equation}
  \epsilon^{\rm GTF}_{\rm po} = {\cal O}(\hbar^1).
  \label{eq:epsgtf}
\end{equation}
That is, the semiclassical energies measured in units of the mean
level spacing are asymptotically accurate independently of the
dimension! This estimate grossly contradicts the ``traditional"
estimate (\ref{eq:epstrad}) and calls for an explanation.

The first point that should be noted is that the order of magnitude
(power of $\hbar$) of the periodic orbit sum, which we considered
above to be ${\cal O}(\hbar^0)$, is only a formal one. Indeed, each
term which is due to a {\it single} periodic orbit is of order ${\cal
O}(\hbar^0)$. However the periodic orbit sum {\it absolutely
diverges}, and at best it is only {\it conditionally convergent}. To
give it a numerical meaning, the periodic orbit sum must therefore be
regularized. This is effectively achieved by truncating the trace
formula or the corresponding spectral $\zeta$ function
\cite{DS92,Bog92b,Kea93,PG95}. However, the truncation cutoff itself
depends on $\hbar$. One can conclude, that the simple minded estimate
(\ref{eq:epsgtf}) given above is at best a lower bound, and the error
introduced by the periodic orbit sum must be re-evaluated with more
care. This point will be dealt with in great detail in the sequel, and
we shall eventually develop a meaningful framework for evaluating the
magnitude of the periodic orbit sum.

The connection and the disparity between the ``traditional" estimate
of the semiclassical error and the one based on the trace formula can
be further illustrated by the following argument.  The periodic orbit
formula is derived from the semiclassical propagator $K_{\rm sc}$
using further approximations \cite{Gut90}. One thus wonders, how can
it be that {\em further} approximations of $K_{\rm sc}$ actually {\em
reduce} the semiclassical error from (\ref{eq:epstrad}) to
(\ref{eq:epsgtf})?  The puzzle is resolved if we recall, that in order
to obtain $\epsilon^{\rm GTF}_{\rm po}$ above we separated the density
of states into a smooth and an oscillating parts, and we required that
the smooth part is accurate enough. To achieve this, we have to go
beyond the leading Weyl's term and to use specialized methods to
calculate the smooth density of states beyond the leading order. These
methods are mostly developed for billiards \cite{BB70,BH76,BH94}. In
any case, to obtain $\epsilon^{\rm GTF}_{\rm po}$ we have added an
{\em additional} information which goes beyond the leading
semiclassical approximation.

A straightforward check of the accuracy of the semiclassical spectrum
using the error measures $\epsilon, \epsilon^{\rm (1)},
\epsilon^{\rm (2)}$ is exceedingly difficult due to the large number of
periodic orbits needed because of the exponential proliferation in
chaotic systems. The few cases where such tests were carried out
involve 2D systems and it was possible to check only the lowest (less
than a hundred) levels (e.g.\ \cite{Sie91,HS92}). The good agreement
between the exact and the semiclassical values confirmed the
expectation that in 2D the semiclassical error is small. In 3D, the
topological entropy is typically much larger \cite{AM96,Pri97}, and
the direct test of the semiclassical spectrum becomes prohibitive.

Facing with this grim reality, we have to introduce alternative error
measures which yield the desired information, but which are more
appropriate for a practical calculation. We construct the measure:
\begin{equation}
  \delta^{\rm (2)}(E) \equiv \left\langle
    \left| N(E) - N_{\rm sc}^{\#}(E) \right|^2
    \right\rangle_{E} \ .
  \label{eq:delta2def}
\end{equation}
As before, the triangular brackets indicate averaging over an energy
interval $\Delta E$ about $E$. We shall now show that $\delta^{\rm
(2)}$ faithfully reflects the deviations between the spectra, and is
closely related to $\epsilon^{\rm (1)}$ and $\epsilon^{\rm
(2)}$. Note, that the following arguments are purely statistical and
apply to every pair of staircase functions.

\begin{figure}[t]
  \begin{center}
    \leavevmode
    \psfig{figure=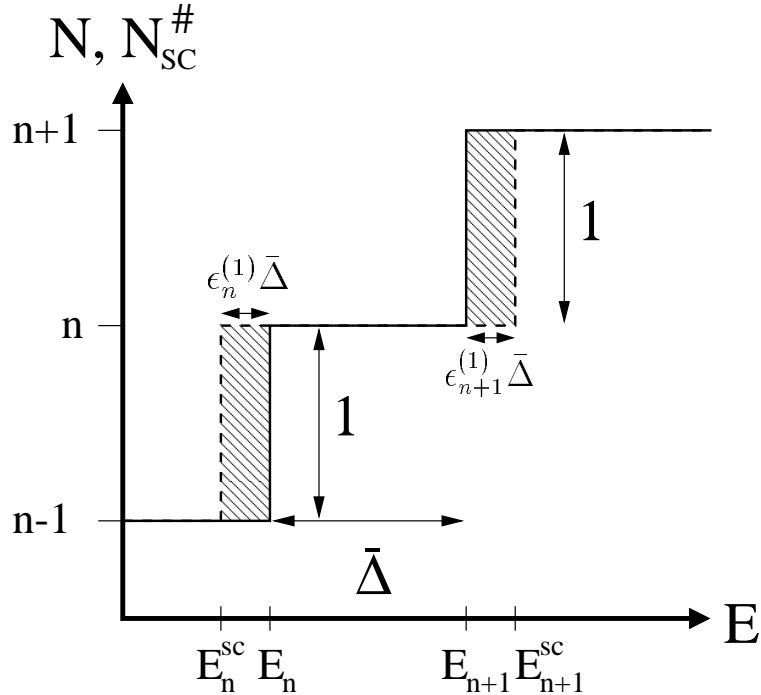,width=10cm}
    \caption{Illustration of $|N(E) - N_{\rm sc}^{\#}(E)|$ for small
      deviations between quantum and semiclassical energies:
      $\epsilon^{\rm (1)} \ll \bar{d}^{-1} \equiv \bar{\Delta}$. The
      quantum staircase $N(E)$ is denoted by the full line and the
      semiclassical staircase $N_{\rm sc}^{\#}(E)$ is denoted by the
      dashed line. The difference is shaded.}
    \label{fig:delta2small}
  \end{center}
\end{figure}

\begin{figure}[t]
  \begin{center}
    \leavevmode
    \psfig{figure=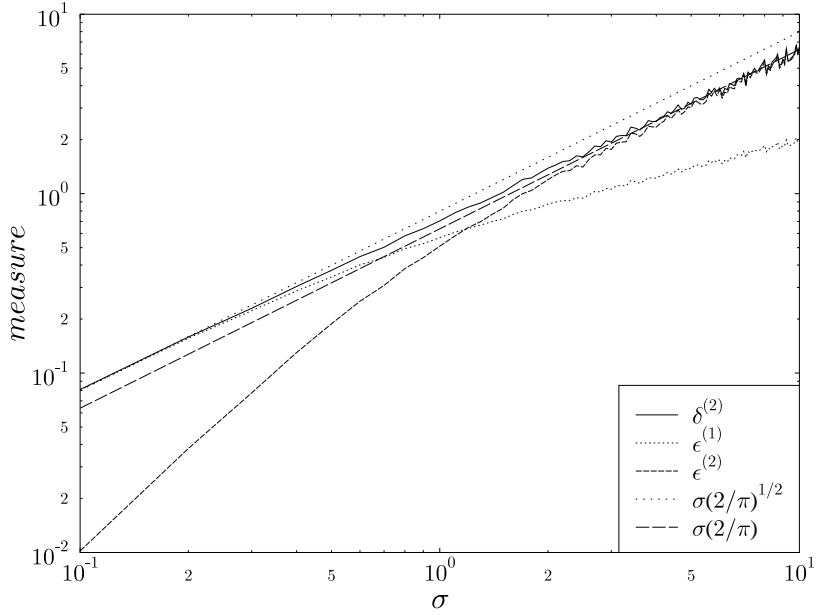,width=12cm}
    \caption{The spectral measures $\epsilon^{\rm (1)}, \epsilon^{\rm
      (2)}, \delta^{\rm (2)}$ and and their asymptotics for a wide range
      of spectral deviations $\sigma$. The random distribution was
      Gaussian, and similar results were obtained also for a uniform
      distribution. Note the logarithmic scale.}
    \label{fig:delta2-eps}
  \end{center}
\end{figure}

\begin{figure}[ht]
  \begin{center}
    \leavevmode
    \psfig{figure=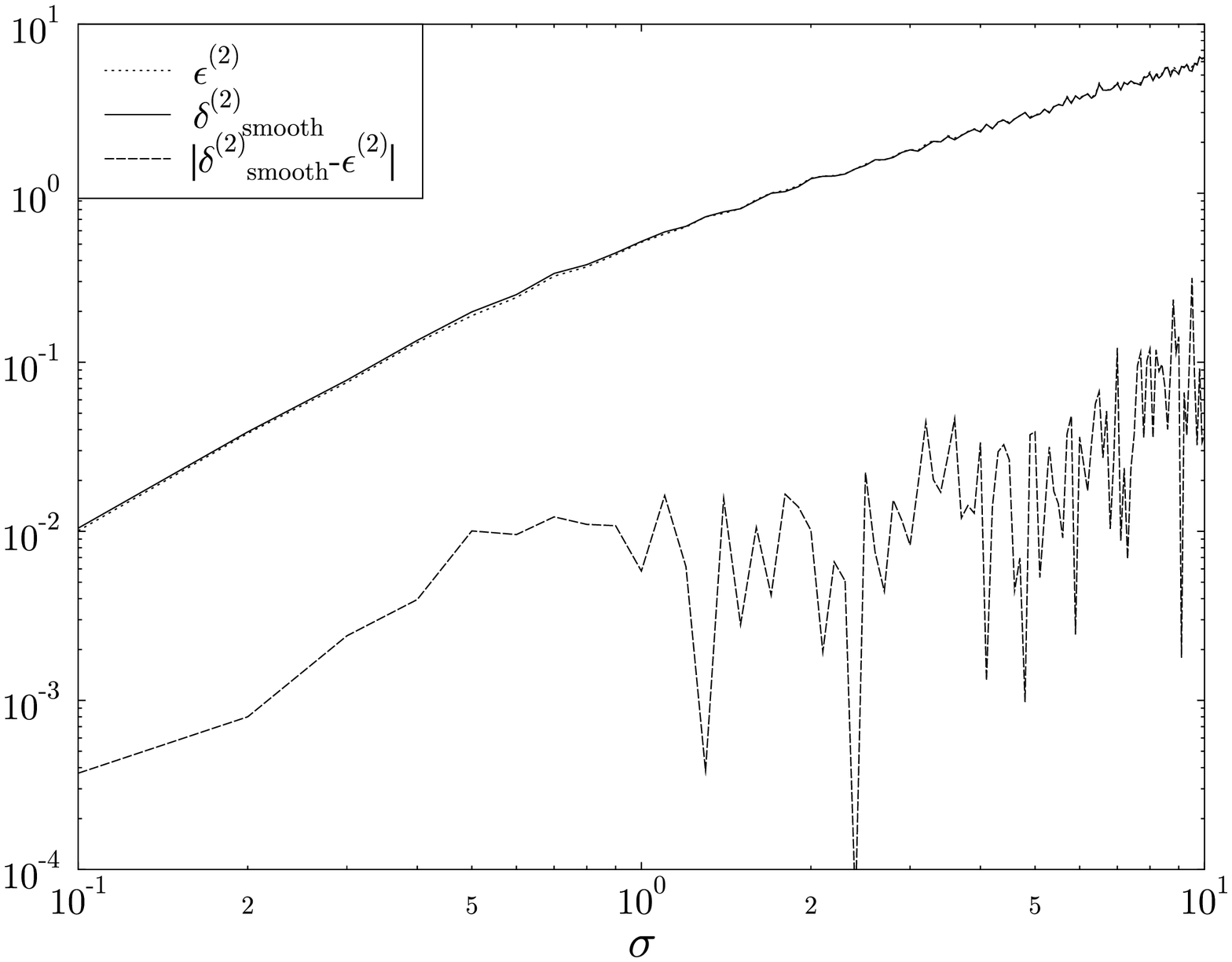,width=12cm}
    \caption{Comparison of the measures $\delta^{\rm (2)}_{\rm smooth}$
      and $\epsilon^{\rm (2)}$ for a wide range of spectral deviations
      $\sigma$ and for Gaussian distribution. The difference is also
      plotted since the curves almost overlap. Note the logarithmic
      scale.}
    \label{fig:delta2smootheps2}
  \end{center}
\end{figure}

Suppose first, that all the differences $E_n^{\rm sc}-E_n$ are smaller
than the mean spacing. Then, $|N - N_{\rm sc}^{\#}|$ is either 0 or 1
(see figure \ref{fig:delta2small}) and hence $|N - N_{\rm sc}^{\#}| =
|N - N_{\rm sc}^{\#}|^2$. Consequently,
\begin{equation}
  \delta^{\rm (2)}(E) \approx \left\langle
    \left| N(E) - N_{\rm sc}^{\#}(E) \right|
   \right\rangle_{E} \; , \; \; \;
   \mbox{small deviations} \, .
\end{equation}
However, the right hand side of the above equation (the fraction of
non--zero contributions) equals $\epsilon^{\rm (1)}$. Thus,
\begin{equation}
  \delta^{\rm (2)} \approx \epsilon^{\rm (1)} \; , \; \; \;
  \mbox{small deviations} \ .
  \label{eq:delta2small}
\end{equation}
If, on the other hand, deviations are much larger than one mean
spacing, the typical horizontal distance $\bar{d}|E-E_n|$ should be
comparable to the vertical distance $|N - N_{\rm sc}^{\#}|$, and hence,
in this limit
\begin{equation}
  \delta^{\rm (2)} \approx \epsilon^{\rm (2)} \; , \; \; \;
  \mbox{large deviations} \, .
  \label{eq:delta2large}
\end{equation}
Therefore, we expect $\delta^{\rm (2)}$ to interpolate between
$\epsilon^{\rm (1)}$ and $\epsilon^{\rm (2)}$ throughout the entire
range of deviations. This behavior is indeed observed in a numerical
test which was performed to check the above expectations. We
considered the unfolded exact spectrum (normalized to unity mean
spacing
\cite{Ber89}) of the 3D Sinai billiard $\{X_n\}$ and created from it a
synthetic spectrum by adding a random variable with $0$ mean and
variance $\sigma^2$:
\begin{equation}
  X^{\sigma}_n = X_n + X_{\rm random}(0, \sigma).
\end{equation}
The $\{X^{\sigma}_{n}\}$ has also a unity mean spacing and it is meant
to imitate a semiclassical spectrum. After sorting
$\{X^{\sigma}_{n}\}$ we calculated the measures $\epsilon^{\rm (1)},
\epsilon^{\rm (2)}$ and $\delta^{\rm (2)}$ as functions of $\sigma$. The
results are shown in figure \ref{fig:delta2-eps}, and they verify the
estimates (\ref{eq:delta2small}, \ref{eq:delta2large}) in the
appropriate limits.
The numerical test reported in figure \ref{fig:delta2smootheps2}
demonstrates another attractive feature of the measure $\delta^{\rm
(2)}$: It is completely equivalent to $\epsilon^{\rm (2)}$ when the
spectral counting functions are replaced by their smooth counterparts,
provided that the smoothing width is of the order of 1 mean level
spacing and the same smoothing is applied to both counting
functions. That is,
\begin{equation}
  \delta^{\rm (2)}_{\rm smooth} \approx \epsilon^{\rm (2)}
  \label{eq:delta2epsilon2}
\end{equation}
for all deviations. (In fact, for small deviations there is a
proportionality factor, but it can be set to 1 if an appropriate
smoothing is used.) In testing the semiclassical accuracy, this kind
of smoothing is essential and will be introduced by truncating the
trace formula at the Heisenberg time $t_{\rm H} \equiv h \bar{d}$.
%
%
These properties of the measure $\delta^{\rm (2)}$, and its complete
equivalence to $\epsilon^{\rm (2)}$ for smooth counting functions,
renders it a most appropriate measure of the semiclassical error.

We now turn to the practical evaluation of $\delta^{\rm (2)}$. To
affect the energy averaging, we choose a positive window function
$w(E'-E)$ which has a width $\Delta E$ near $E$ and is normalized by
$\int_{-\infty}^{+\infty} {\rm d}E' w^2 (E') = 1$. It falls off
sufficiently rapidly so that all the expressions which follow are well
behaved. Construct the following counting functions that have an
effective support on an interval of size $\Delta E$ about $E$:
\begin{eqnarray}
  \hat{N}(E'; E)
   & \equiv &
   w(E'{-}E) N(E') \\
   \hat{N}_{\rm sc}^{\#}(E'; E)
   & \equiv &
   w(E'{-}E) N_{\rm sc}^{\#}(E') \, .
\end{eqnarray}
At this stage $\hat{N}$ and $\hat{N}_{\rm sc}^{\#}$ are still sharp
staircases, and we note that the multiplication with~$w$ preserves the
sharpness of the stairs (it is not a convolution!). We now explicitly
construct $\delta^{\rm (2)}(E)$ as:
\begin{eqnarray}
  \delta^{\rm (2)}(E)
  & = &
  \int_{-\infty}^{+\infty} {\rm d}E' \,
  \left| \hat{N}(E'; E) - \hat{N}_{\rm sc}^{\#}(E'; E) \right|^2
  \nonumber \\
  & = &
  \int_{-\infty}^{+\infty} {\rm d}E' \,
  \left| N(E') - N_{\rm sc}^{\#}(E') \right|^2
   w^2 (E'-E).
   \label{eq:delta2int}
\end{eqnarray}
To construct $\delta^{\rm (2)}_{\rm smooth}$ we need to smooth $N,
N_{\rm sc}^{\#}$ over a scale of order of one mean spacing. This can
be done e.g.\ by replacing the sharp stairs by error functions. As for
$N_{\rm sc}^{\#}$, we prefer to simply replace it with the original
$N_{\rm sc}$, which we assume to be smooth over one mean spacing. That
is, we suppose that $N_{\rm sc}$ contains periodic orbits up to
Heisenberg time. Hence,
\begin{equation}
  \delta^{\rm (2)}_{\rm smooth}(E) =
  \int_{-\infty}^{+\infty} {\rm d}E' \,
  \left| N^{\rm smooth}(E') - N_{\rm sc}(E') \right|^2
   w^2 (E'-E)  .
  \label{eq:delta2smoothint}
\end{equation}
A comment is in order here. Strictly speaking, to satisfy
(\ref{eq:delta2epsilon2}) we need to apply the same smoothing to $N$
and to $N_{\rm sc}^{\#}$, and in general $N_{\rm sc}^{\#, {\rm smooth}} \ne
N_{\rm sc}$, but there are differences of order 1 between the two
functions. However, since our goal is to determine whether the
semiclassical error remains finite or diverges in the semiclassical
limit $\hbar \rightarrow 0$, we disregard such inaccuracies of order
1. If more accurate error measure is needed, then more care should be
practiced in this and in the following steps.

Applying Parseval's theorem to (\ref{eq:delta2smoothint}) we get:
\begin{equation}
  \delta^{\rm (2)}_{\rm smooth}(E) =
  \frac{1}{\hbar} \int_{-\infty}^{+\infty} {\rm d}t \,
  \left| \hat{D}(t; E) - \hat{D}_{\rm sc}(t; E) \right|^2
  \label{eq:delta2ls}
\end{equation}
where
\begin{eqnarray}
  \hat{D}(t; E)
  & \equiv &
  \frac{1}{\sqrt{2 \pi}} \int_{-\infty}^{+\infty} {\rm d}E' \,
  \hat{N}^{\rm smooth}(E'; E) \exp(i E' t / \hbar) \\
  \hat{D}_{\rm sc}(t; E)
  & \equiv &
  \frac{1}{\sqrt{2 \pi}} \int_{-\infty}^{+\infty} {\rm d}E' \,
  \hat{N}_{\rm sc}(E'; E) \exp(i E' t / \hbar).
\end{eqnarray}
We shall refer to $\hat{D}, \hat{D}_{\rm sc}$ as the (regularized)
quantal and semiclassical time spectra, respectively. This name can be
justified by invoking the Gutzwiller trace formula and expressing the
semiclassical counting function as a mean part plus a sum over
periodic orbits. We have:
\begin{equation}
  N_{\rm sc}(E) =
  \bar{N}(E) +
  \sum_{\rm po} \frac{\hbar A_j(E)}{T_j(E)}
  \sin [ S_j(E) / \hbar - \nu_j \pi / 2 ] \ ,
  \label{eq:gtf}
\end{equation}
where $A_j = T_j / (\pi \hbar r_j \sqrt{|I - M_j|})$ is the
semiclassical amplitude of the j'th periodic orbit, and $T_j, S_j,
\nu_j, M_j, r_j$ are its period, action, Maslov index, monodromy and
repetition index, respectively. Then, the corresponding time spectrum
reads:
\begin{eqnarray}
  \hat{D}_{\rm sc}(t; E)
  & \approx &
  \bar{D}(t; E) 
  \label{eq:scltspt} \\
  & + &
  \frac {1}{2i}\sum_{\rm po}
  \frac{\hbar A_j(E)}{T_j(E)}
  \left\{{\rm e}^{(i/\hbar)[E t + S_j(E)]} 
    \hat{w}([t + T_j(E)]/\hbar) - 
  \right. \nonumber \\
  & &
  \left. \; \; \; \; \; \; \; \; \; \; \; \; \; \; \; \; \; \;
         \; \; \; \; \;
  {\rm e}^{(i/\hbar)[E t - S_j(E)]} 
    \hat{w}([t - T_j(E)]/\hbar)
  \right\} .
  \nonumber
\end{eqnarray}
In the above, the Fourier transform of $w$ is denoted by $\hat{w}$. It
is a localized function of $t$ whose width is $\Delta t \approx \hbar
/ \Delta E$. The sum over the periodic orbits in $D_{\rm sc}$
therefore produces sharp peaks centered at times that correspond to
the periods $T_j$, hence the name ``time spectrum''. The term
$\bar{D}$ corresponds to the smooth part and thus is sharply peaked
near $t=0$. To obtain (\ref{eq:scltspt}) we expanded the actions near
$E$ to first order: $S_j(E') \approx S_j(E) + (E'-E)T_j(E)$. We note
in passing, that this approximate expansion of $S_j$ can be avoided
altogether if one performs the Fourier transform over $\hbar^{-1}$
rather than over the energy. This way, an action spectrum will emerge,
but also here the action resolution will be finite, because the range
of $\hbar^{-1}$ should be limited to the range where $\bar d(E;\hbar)$
is approximately constant. It turns out therefore, that the two
approaches are essentially equivalent, and for billiards they are
identical.

The manipulations done thus far were purely formal, and did not
manifestly circumvent the difficult task of evaluating $\delta^{\rm
(2)}_{\rm smooth}$. However, the introduction of the time spectra and
the formula (\ref{eq:delta2ls}) put us in a better position than with
the original expression (\ref{eq:delta2int}). The advantages of using
the time spectra in the present context are the following:
\begin{itemize}

\item The semiclassical time spectrum $\hat{D}_{\rm sc}(t;E)$ is absolutely
  convergent for all times (as long as the window function~$w$ is well
  behaved, e.g.\ it is a Gaussian). This statement is correct even if
  the sum (\ref{eq:scltspt}) extends over the entire set of periodic
  orbits! This is in contrast with the trace formula expression for
  $N_{\rm sc}$ (and therefore $\hat{N}_{\rm sc}$) which is absolutely
  divergent if all of the periodic orbits are included.

\item Time scale separation: As we noted above, the time spectrum
  is  peaked at times that correspond to periods of the classical
  periodic orbits. This allows us to distinguish between various
  qualitatively different types of contributions to
  $\delta^{\rm (2)}_{\rm smooth}$.

\end{itemize}
We shall now pursue the separation of the time scales in detail. We
first note, that due to $\hat{N}, \hat{N}_{\rm sc}$ being real, there
is a $t \leftrightarrow (-t)$ symmetry in (\ref{eq:delta2ls}) and
therefore $\delta^{\rm (2)} = 2 \int_0^\infty \cdots$. We now divide
the time axis into four intervals :
\begin{description}

\item{$\bf 0 \leq t \leq \Delta t$}: The shortest time scale in our
  problem is $\Delta t = \hbar / \Delta E$. The contributions to this
  time interval are due to the differences between the exact and the
  semiclassical {\em mean} densities of states. This is an important
  observation, since it allows us to distinguish between the two
  sources of semiclassical error --- the error that emerges from the
  mean densities and the error that originates from the fluctuating
  part (periodic orbits). Since we are interested only in the
  semiclassical error that results from the fluctuating part of the
  spectral density, we shall ignore this regime in the following.

\item[$\bf \Delta t \leq t \leq t_{\rm erg}$]: This is the non--universal
  regime \cite{Ber89}, in which periodic orbits are still sparse, and
  cannot be characterized statistically in a significant fashion.  The
  ``ergodic'' time scale $t_{\rm erg}$ is purely classical and is
  independent of $\hbar$.

\item{$\bf t_{\rm erg} \leq t \leq t_{\rm H} $}: In this time regime
  periodic orbits are already in the universal regime and are dense
  enough to justify a statistical approach to their proliferation and
  stability. The upper limit of this interval is the Heisenberg time
  $t_{\rm H} = h\bar{d}(E)$, which is the time that is needed to
  resolve the quantum (discrete) nature of a wavepacket with energy
  concentrated near $E$. The Heisenberg time is ``quantum'' in the
  sense that it is dependent of $\hbar$: $t_{\rm H} = {\cal
  O}(\hbar^{1-d})$.

\item{$\bf t_{\rm H} \leq t < \infty$}: This is the interval of ``long''
  orbits which is effectively truncated from the integration as a
  result of introducing a smoothing of the quantal and semiclassical
  counting functions, with a smoothing scale of the order of a mean
  level spacing.

\end{description}
Dividing the integral (\ref{eq:delta2ls}) according to the above time
intervals, we can rewrite $\delta^{\rm (2)}_{\rm smooth}$:
\begin{eqnarray}
  \delta^{\rm (2)}_{\rm smooth}(E)
  & = &
  \left( \int_{\Delta t}^{t_{\rm erg}} +
  \int_{t_{\rm erg}}^{t_{\rm H}} + \int_{t_{\rm H}}^{\infty} \right)
  \frac{2 {\rm d}t}{\hbar} \,
  \left| \hat{D}(t; E) - \hat{D}_{\rm sc}(t; E) \right|^2
  \nonumber \\
  & \equiv &
  \delta^{\rm (2)}_{\rm short} +
  \delta^{\rm (2)}_{\rm m} +
  \delta^{\rm (2)}_{\rm long} \; .
  \label{eq:delta2split}
\end{eqnarray}
As explained above, $\delta^{\rm (2)}_{\rm long}$ can be ignored due
to smoothing on the scale of a mean level spacing. The integral
$\delta^{\rm (2)}_{\rm short}$ is to be neglected for the following
reason.  The integral extends over a time interval which is finite and
independent of $\hbar$, and therefore it contains a fixed number of
periodic orbits contributions. The semiclassical approximation
provides, for each individual contribution, the leading order in
$\hbar$, and therefore \cite{AM77} we should expect:
\begin{equation}
  \delta^{\rm (2)}_{\rm short} \longrightarrow 0 \; \; \;
  \mbox{as } \; \hbar \longrightarrow 0.
  \label{eq:delta2-short}
\end{equation}
The purpose of this work is to check whether the semiclassical error
is finite or divergent as $\hbar \longrightarrow 0$, and to study if
the rate of divergence depends on dimensionality. Equation
(\ref{eq:delta2-short}) implies that $\delta^{\rm (2)}_{\rm short}$
cannot affect $\delta^{\rm (2)}$ in the semiclassical limit and we
shall neglect it in the following.

We thus remain with a lower bound for our measure:
\begin{equation}
  \delta^{\rm (2)}_{\rm smooth} \approx 
  \delta^{\rm (2)}_{\rm m}
\end{equation}
which is going to be our object of interest from now on.

The fact that $t_{\rm H}$ is extremely large on the classical scale
renders the calculation of all the periodic orbits with periods less
than $t_{\rm H}$ an impossible task. However, sums over periodic
orbits when the period is longer than $t_{\rm erg}$ tend to meaningful
limits, and hence, we would like to recast the expression for
$\delta^{\rm (2)}_{\rm m}$ in the following way.  Write $\delta^{\rm
(2)}_{\rm m}$ as:
\begin{eqnarray}
  \delta^{\rm (2)}_{\rm m}
  & = &
  \frac{2}{\hbar} \int_{t_{\rm erg}}^{t_{\rm H}} {\rm d}t \,
  \left\langle \left| \hat{D}(t) - \hat{D}_{\rm sc}(t) \right|^2
  \right\rangle_{t} \\
  & = &
  \frac{2}{\hbar} \int_{t_{\rm erg}}^{t_{\rm H}} {\rm d}t \,
  \left\langle \left| \hat{D}(t) \right|^2 \right\rangle_{t}
  \times \left[
    \frac{\left\langle \left| \hat{D}(t) - \hat{D}_{\rm sc}(t)
        \right|^2 \right\rangle_{t}}
    {\left\langle \left| \hat{D}(t) \right|^2 \right\rangle_{t}}
  \right]
  \label{eq:ct} \\
  & \equiv &
  \frac{2}{\hbar} \int_{t_{\rm erg}}^{t_{\rm H}} {\rm d}t \,
  \left\langle \left| \hat{D}(t) \right|^2
  \right\rangle_{t} \times C(t) \\
  & = &
  \int_{t_{\rm erg}}^{t_{\rm H}} \mbox{\large envelope} \times
  \mbox{\large correlation} \nonumber \,
\end{eqnarray}
where the parametric dependence on $E$ was omitted for brevity. The
smoothing over $t$ is explicitly indicated to emphasize that one may
use a statistical interpretation of the terms of the integrand. This
is so because in this domain, the density of periodic orbits is so
large, that within a time interval of width $\hbar/\Delta E$ there are
exponentially many orbits whose contributions are averaged due to the
finite resolution.

We note now that we can use the following relation between the time
spectrum and the spectral form factor $K(\tau)$:
\begin{equation}
  \frac{\left\langle \left| \hat{D}(t) \right|^2
    \right\rangle_{t}}{\hbar}
  \, {\rm d}t =
  \frac{K(\tau)}{4 \pi^2 \tau^2} \, {\rm d}\tau
  \label{eq:dtktau}
\end{equation}
where $\tau \equiv t/t_{\rm H}$ is the scaled time. The above form
factor is smoothed according to the window function ~$w$. Hence:
\begin{equation}
  \delta^{\rm (2)}_{\rm smooth} \approx \frac{1}{2 \pi^2}
  \int_{\tau_{\rm erg}}^{1} {\rm d}\tau \,
  \frac{K(\tau) C(\tau)}{\tau^2} \, .
\end{equation}
For generic chaotic systems we expect that $K(\tau)$ agrees with the
results of RMT in the universal regime $\tau > \tau_{\rm erg}$
\cite{BGS84,Ber85,Ber89}, and therefore
\begin{equation}
  K(\tau) \approx g \tau \;\;\;
  \mbox{for } \tau_{\rm erg} < \tau \leq 1 ,
\end{equation}
where $g=1$ for systems which violate time reversal symmetry, and
$g=2$ if time reversal symmetry is respected. This implies that the
evaluation of $\delta^{\rm (2)}_{\rm smooth}$ reduces to
\begin{equation}
  \delta^{\rm (2)}_{\rm smooth} \approx \frac{g}{2 \pi^2}
  \int_{\tau_{\rm erg}}^{1} {\rm d}\tau \, \frac{C(\tau)}{\tau}.
  \label{eq:delta2final}
\end{equation}
The dependence on $\hbar$ in this expression comes from
the lower integration limit which is proportional to $\hbar^{d-1}$
as well as from the implicit dependence of the function $C$ on $\hbar$.

Formula (\ref{eq:delta2final}) is our main theoretical
result. However, we do not know how to evaluate the correlation
function $C(\tau)$ from first principles. The knowledge of the $\hbar
$ corrections to each of the terms in the semiclassical time spectrum
is not sufficient since the resulting series which ought to be summed
is not absolutely convergent (see detailed discussion in section
(\ref{sec:discussion})). Therefore we have to recourse to a numerical
analysis, which will be described in the next section. The numerical
approach requires one further approximation, which is imposed by the
fact that the number of periodic orbits with $t<t_{\rm H}$ is
prohibitively large. We had to limit the data base of periodic orbit
to the domain $t< t_{\rm cpu}$ with $t_{\rm erg} \ll t_{\rm cpu}
\ll t_{\rm H}$. The time $t_{\rm cpu}$ has no physical origin, 
and it represents only the limits of our computational
resources. Using the available numerical data we were able to compute
$C(t)$ numerically for all $ t_{\rm erg}<t<t_{\rm cpu}$ and we then
{\em extrapolated} it to the entire domain of interest. We consider
this extrapolation procedure to be the main source of
uncertainty. However, since the extrapolation is carried out in the
{\em universal} regime, it should be valid if there are no other time
scales between $t_{\rm erg}$ and $t_{\rm H}$.

\section{Numerical results}
\label{sec:sinai-numerics}
%
We used the formalism and definitions presented above to check the
accuracy of the semiclassical spectra of the 2D and 3D Sinai
billiards. The most important ingredient in this numerical study is
that we could apply the {\em same} analysis to the two systems, and by
comparing them to give a reliable answer to the main question posed in
this work, namely, how does the semiclassical accuracy depend on
dimensionality.

The classical dynamics in billiards depends trivially on the energy
(velocity), and therefore the relevant parameter is the length rather
than the period of the periodic orbits. Because of the same reason,
the quantum wavenumbers $k_n \equiv \sqrt{2 m E_n}/\hbar$ are the
relevant variables in the quantum description. From now on we shall
use the variables $(l, k)$ instead of $(t, E)$, and use ``length
spectra'' rather than ``time spectra''. The semiclassical limit is
obtained for $k \rightarrow \infty$ and ${\cal O}(\hbar)$ is
equivalent to ${\cal O}(k^{-1})$. Note also that for a billiard
$\bar{N}(k) \approx A k^{d}$ where $A$ is a proportional to the
billiard's volume.

The numerical work is based on the quantum spectra and on the
classical periodic orbits which were calculated by Schanz and
Smilansky \cite{SS95,Sch96a} for the 2D billiard, and by Primack and
Smilansky \cite{PS95,Pri97} for the 3D billiard. The numerical methods
and the checks performed to ensure that the quantum and the classical
data bases are accurate, complete and immaculate are discussed in the
papers cited above.

\begin{figure}[t]
  \begin{center}
    \leavevmode
    \begin{tabular}{ccc}
      \psfig{figure=FIGURES/sb2d.eps,width=5cm} &
      \hspace{2cm} &
      \psfig{figure=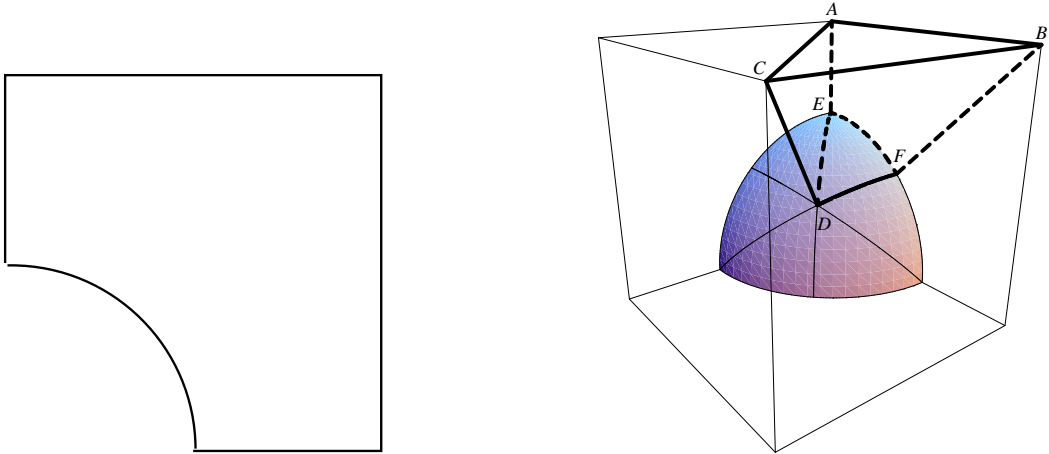,width=6cm} 
    \end{tabular}
    \caption{Left: The quarter desymmetrized 2D Sinai billiard. 
             Right: The fully ($1/48$) desymmetrized 3D Sinai 
             billiard (bold lines).}
    \label{fig:sb2d-sb3d}
  \end{center}
\end{figure}

We start with the 2D Sinai billiard, which is the free space between a
square of edge $L$ and an inscribed disc of radius $R$, with $2R <
L$. In our case we used $L=1$ and $R=0.25$ and considered the quarter
desymmetrized billiard (see figure \ref{fig:sb2d-sb3d}) with Dirichlet
boundary conditions for the quantum calculations. The quantal data
base consisted of the lowest 27645 eigenvalues in the range $0 < k
<1320$, with eigenstates which are either symmetric or antisymmetric
with respect to reflection on the main diagonal.  The classical data
base consisted of the shortest 20273 periodic orbits (including time
reversal, reflection symmetries and repetitions) in the length range
$0 < l < 5$. For each orbit, the length, the stability determinant and
the reflection phase were recorded.

\begin{figure}[t]
  \begin{center}
    \leavevmode
    \psfig{figure=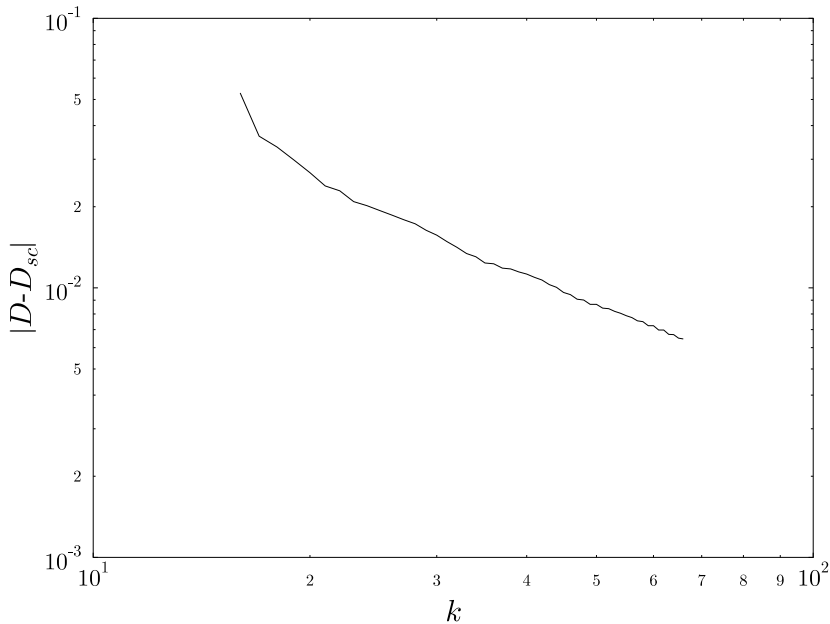,width=12cm}
    \caption{The absolute difference between the quantal and the
             semiclassical (Gutzwiller) length spectra for the 2D
             Sinai billiard at $l=0.5$. This length corresponds to the
             shortest unstable periodic orbit. The average log--log
             slope is about $-1.1$, indicating approximately $k^{-1}$
             decay. The data were averaged with a Gaussian window.}
    \label{fig:sb2d-qm-sc}
  \end{center}
\end{figure}

We begin the numerical analysis of the 2D Sinai billiard by
numerically demonstrating the correctness of equation
(\ref{eq:delta2-short}). That is, that for each {\em individual}
periodic orbit, the semiclassical error indeed vanishes in the
semiclassical limit. In figure \ref{fig:sb2d-qm-sc} we plot $|D -
D_{\rm sc}|$ for $l=0.5$ as a function of $k$. This length corresponds
to the shortest periodic orbit, that is, the one that runs along the
edges that connect the circle with the outer square. For $D_{\rm sc}$
we used the Gutzwiller trace formula. As is clearly seen from the
figure, the quantal--semiclassical difference indeed vanishes
(approximately as $k^{-1}$), in accordance with
(\ref{eq:delta2-short}). We emphasize again, that this behavior does
not imply that $\delta^{\rm (2)}$ vanishes in the semiclassical limit,
since the number of terms depends on $k$. It implies {\em only} that
$\delta^{\rm (2)}_{\rm short}$ vanishes in the limit, since it
consists of a fixed and finite number of periodic orbit
contributions. We should also comment that (non--generic) penumbra
corrections to individual grazing orbits introduce errors which are of
order $k^{-\gamma}$ with $0< \gamma < 1$ \cite
{PSSU96,PSSU97}. However, since the definition of ``grazing'' is in
itself $k$ dependent, one can safely neglect penumbra corrections in
estimating the large $k$ behavior of $\delta^{\rm (2)}_{\rm short}$.

\begin{figure}[t]
  \begin{center}
    \leavevmode
    \psfig{figure=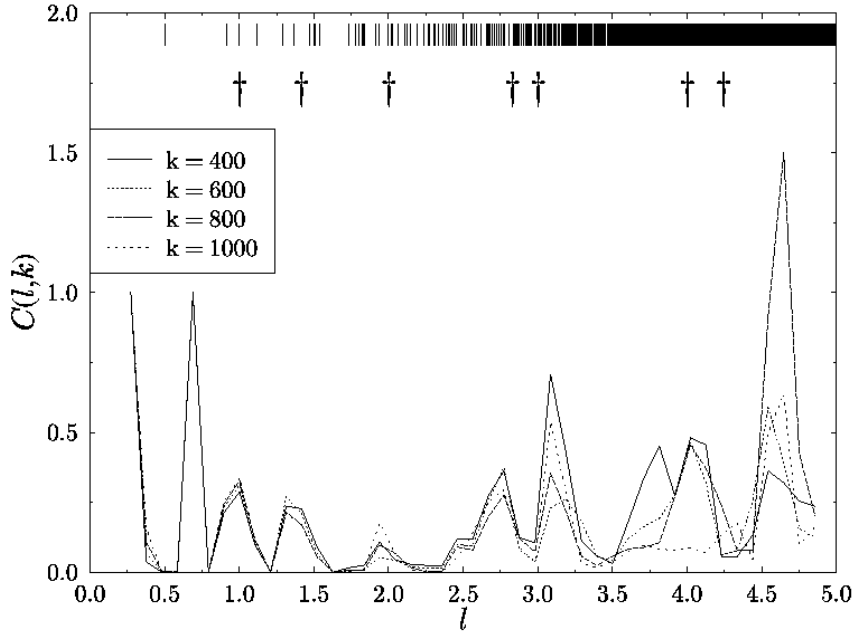,width=12cm}

    \caption{The functions $C(l; k)$ for quarter 2D Sinai billiard
      $L=1, R=0.25$ with Dirichlet boundary conditions. The window
      $w(k'-k)$ was taken to be a Gaussian with standard deviation
      $\sigma = 60$. We averaged $C(l; k)$ over $l$-intervals of
      $\approx 0.2$ in accordance with (\protect\ref{eq:ct}) to avoid
      sharp peaks due to small denominators. The averaging, however,
      is fine enough not to wash out all of the features of $C(l; k)$.
      The vertical bars indicate the locations of primitive periodic
      orbits, and the daggers indicate the locations of the
      bouncing--ball families.}

    \label{fig:dcxavg.2}
  \end{center}
\end{figure}

\begin{figure}[ht]
  \begin{center}
    \leavevmode
    \psfig{figure=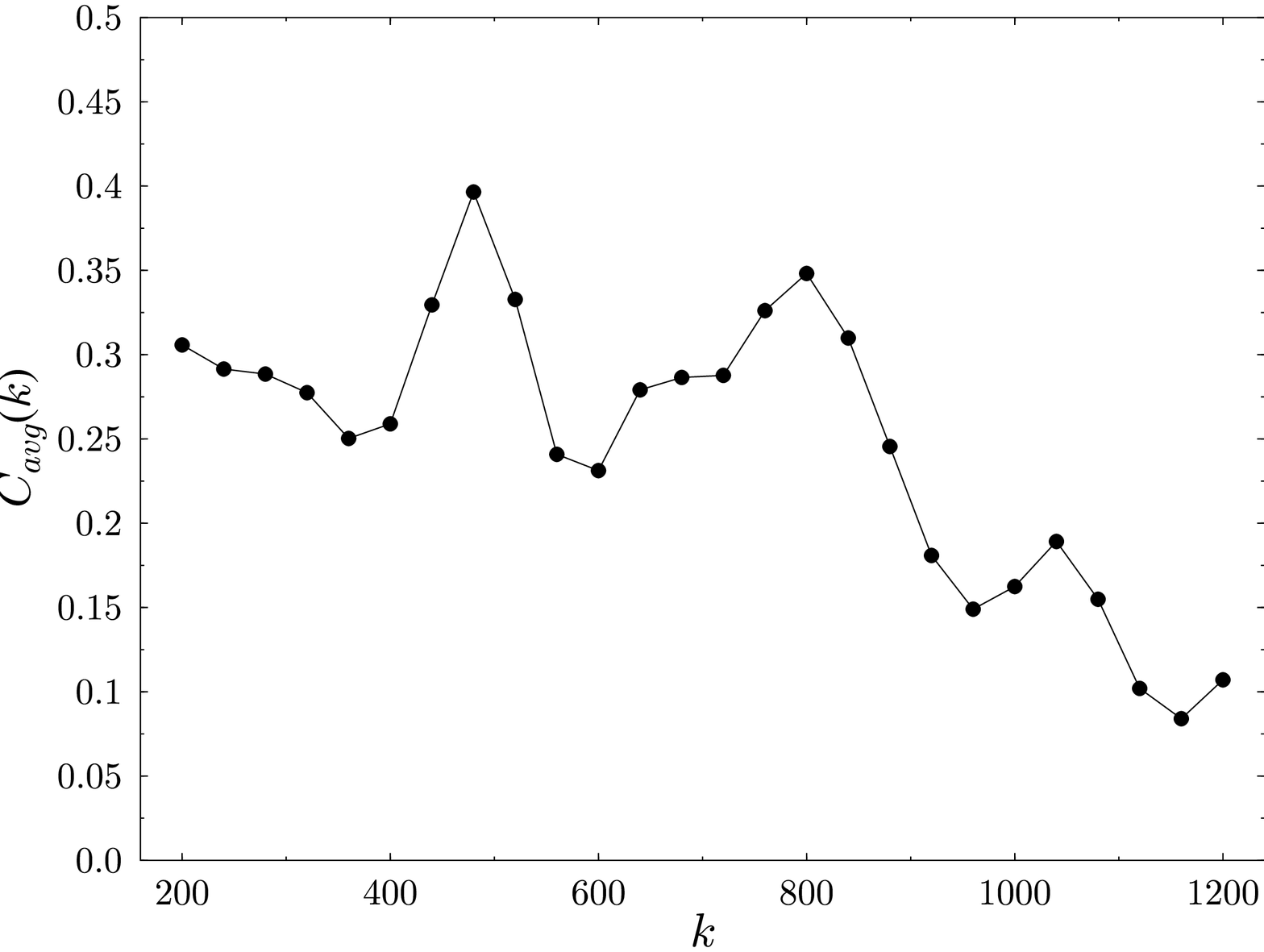,width=12cm}
    \caption{Averaging in $l$ of $C(l; k)$ for 2D Sinai billiard
      as a function of $k$.}
    \label{fig:dcavg.2}
  \end{center}
\end{figure}

We now turn to the main body of the analysis, which is the evaluation
of $\delta^{\rm (2)}_{\rm m}$ for the 2D Sinai billiard. Based on the
available data sets, we plot in figure \ref{fig:dcxavg.2} the function
$C(l; k)$ in the interval $2.5 < l < 5$ for various values of $k$. One
can observe, that as a function of $l$ the functions $C(l; k)$
fluctuate in the interval for which numerical data were available,
without exhibiting any systematic mean trend to increase or to
decrease. We therefore approximate $C(l; k)$ by
\begin{equation}
  C(l; k) \approx {\rm const} \cdot f(k) \equiv C_{\rm avg}(k).
\end{equation}
According to the discussion in section \ref{sec:accuracy-measures} we
extrapolate this formula in $l$ up to the Heisenberg length $L_{\rm H}
= 2 \pi \bar{d}(k)$ and using (\ref{eq:delta2final}) we obtain:
\begin{equation}
  \delta^{\rm (2), 2D}_{\rm smooth} = 
  \frac{C_{\rm avg}(k)}{2 \pi^2} \ln(L_{\rm H} /
  L_{\rm erg}) = C_{\rm avg}(k) \ {\cal O}(\ln k).
\end{equation}
The last equality is due to $L_{\rm H} = {\cal O}(k^{d-1})$. To
evaluate $C_{\rm avg}(k)$ we averaged $C(l; k)$ over the interval
$L_{\rm erg} = 3.5 < l < 5 = L_{\rm cpu}$ and the results are shown in
figure \ref{fig:dcavg.2}. We choose $L_{\rm erg} = 3.5$ because the
density of periodic orbits is already large for this length (see
figure \ref{fig:dcxavg.2}) and we expect universal behavior of the periodic
orbits. (For the Sinai billiard described by flow the approach to the
invariant measure is algebraic rather than exponential
\cite{FS95,DA96}, and thus one cannot have a well-defined $L_{\rm erg}$.
An any rate, the specific choice of $L_{\rm erg}$ did not affect the
results in any appreciable way.) Inspecting $C_{\rm avg}(k)$, it is
difficult to arrive at firm conclusions, since it seems to fluctuate
around a constant value up to $k \approx 900$ and then to decline. If
we approximate $C_{\rm avg}(k)$ by a constant, we get a
``pessimistic'' value of $\delta^{\rm (2)}$:
\begin{equation}
  \delta^{\rm (2), 2D}_{\rm smooth}(k) =
  {\cal O}(\ln k) = {\cal O}(\ln \hbar) \; \; \;
  \mbox{``pessimistic''}
  \label{eq:delta2-pessimistic}
\end{equation}
while if we assume that $C_{\rm avg}(k)$ decays as a power-law,
$C_{\rm avg}(k) = k^{-\beta}, \beta > 0$, then
\begin{equation}
  \delta^{\rm (2), 2D}_{\rm smooth}(k) = 
  {\cal O}(k^{-\beta} \ln k)
  \longrightarrow 0 \; \; \;
  \mbox{``optimistic''} \ .
\end{equation}
Collecting the two bounds we get:
\begin{equation}
  {\cal O}(k^{-\beta} \ln k) \leq
  \delta^{\rm (2), 2D}_{\rm smooth}(k) \leq
  {\cal O}(\ln k) \, .
  \label{eq:delta2-2d}
\end{equation}
Our estimates for the 2D Sinai billiard can be summarized by saying
that the semiclassical error diverges no worse than logarithmically
(meaning, very mildly). It may well happen that the semiclassical
error is constant or even vanishes in the semiclassical limit.  To
reach a conclusive answer one should invest exponentially larger
amount of numerical work.

\begin{figure}[t]
  \begin{center}
    \leavevmode
    \psfig{figure=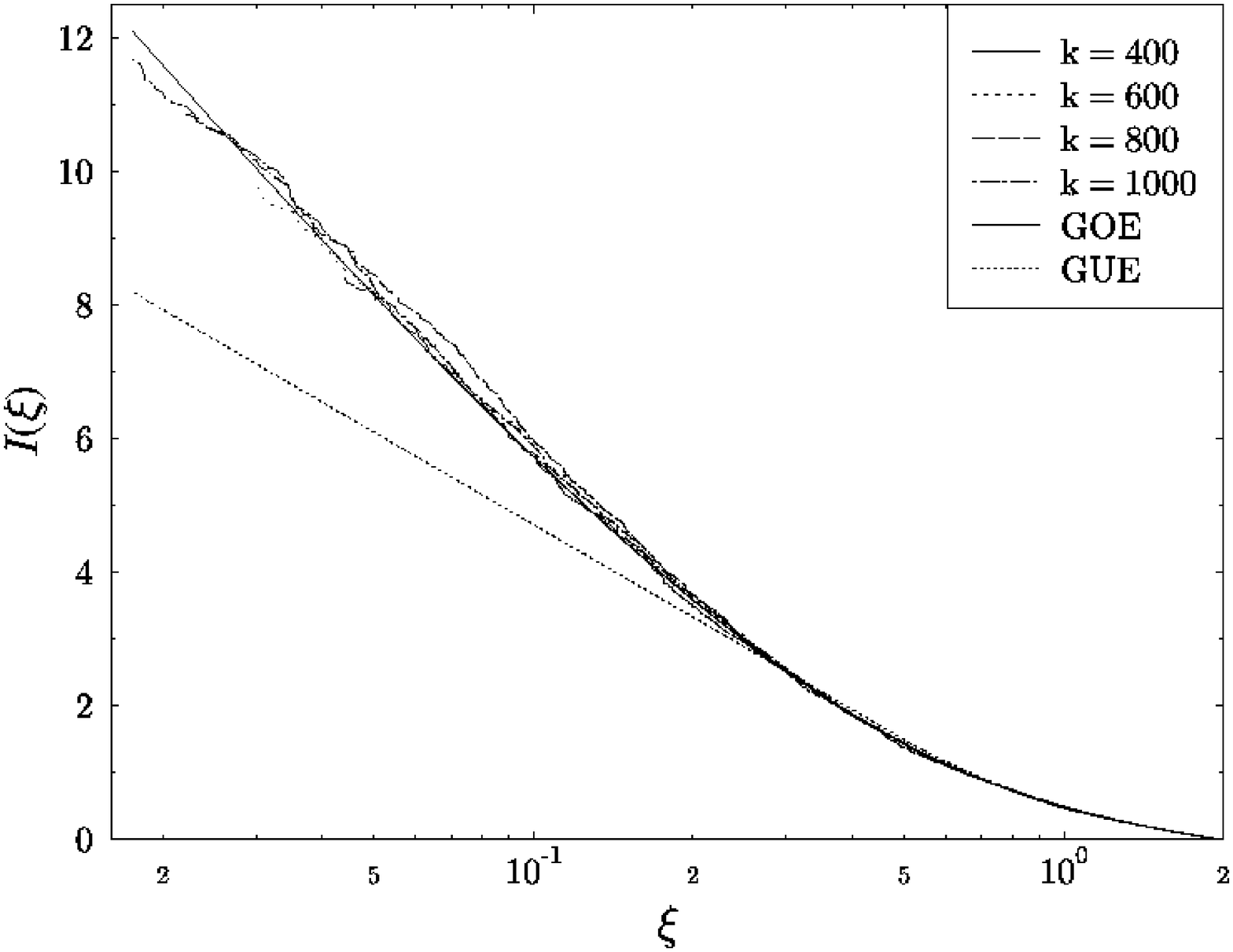,width=12cm}
    \caption{Verification of equation (\protect\ref{eq:dxkxi}) for the
      quarter 2D Sinai billiard. We plot $I(\xi) \equiv \int_{\xi}^{2}
      {\rm d}\xi' \ K(\xi')/\xi'\,^{2}$ and compare the quantum data
      with RMT. The minimal $\xi$ corresponds to $L_{\rm
      erg}=3.5$. The integration is done for smoothing, and we fix the
      {\em upper} limit to avoid biases due to non--universal
      regime. Note the logarithmic scale.}
    \label{fig:dikt.2}
  \end{center}
\end{figure}

\begin{figure}[ht]
  \begin{center}
    \leavevmode
    \psfig{figure=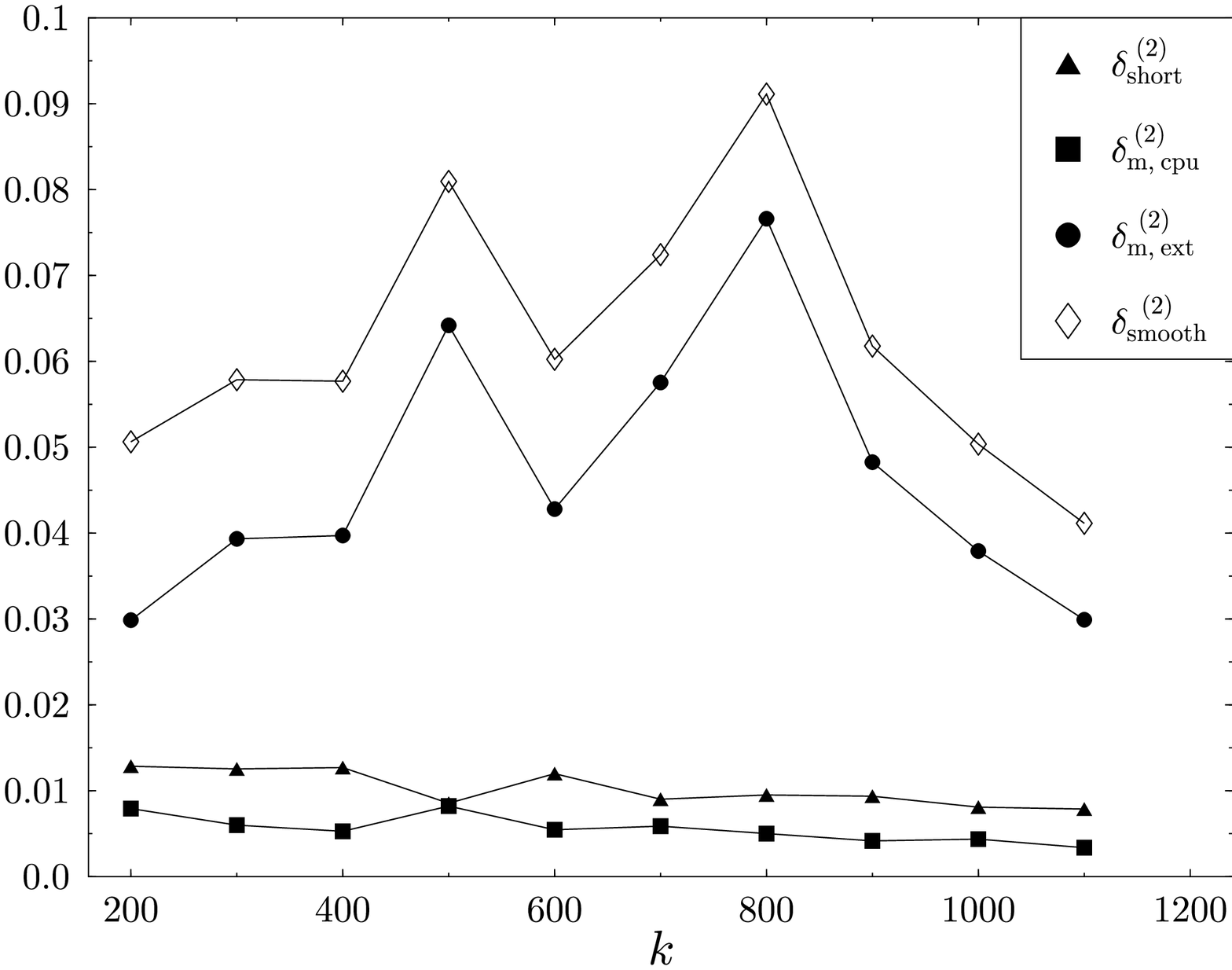,width=12cm} 
    \caption{The numerical values of $\delta^{\rm (2)}_{\rm smooth}$
    for the quarter 2D Sinai billiard. We included also the
    contribution $\delta^{\rm (2)}_{\rm short}$ of the non--universal
    regime. The contributions from the time interval $t_{\rm erg} \leq
    t \leq t_{\rm cpu}$ are contained in $\delta^{\rm (2)}_{\rm m,
    cpu}$, and $\delta^{\rm (2)}_{\rm m, ext}$ is the extrapolated
    value for $t_{\rm cpu} \leq t \leq t_{\rm H}$ (refer to
    (\protect\ref{eq:delta2split}) and to the end of section
    \protect\ref{sec:accuracy-measures}).}
    \label{fig:ddelta2.2}
  \end{center}
\end{figure}

There are a few comments in order here. Firstly, the quarter
desymmetrization of the 2D Sinai billiard does not exhaust its
symmetry group, and in fact, a reflection symmetry around the diagonal
of the square remains. This means, that the spectrum of the quarter 2D
Sinai billiard is composed of two independent spectra, which differ by
their parity with respect to the diagonal. If we assume that the
semiclassical deviations of the two spectra are not correlated, the
above measure is the sum of the two independent measures. It is
plausible to assume also that both spectra have roughly the same
semiclassical deviation, and thus $\delta^{\rm (2), 2D}_{\rm smooth}$
is twice the semiclassical deviation of each of the spectra. Secondly,
we recall that the 2D Sinai billiard contains ``bouncing--ball''
families of neutrally stable periodic orbits
\cite{Ber81,SSCL93,SS95}. We have subtracted their leading-order
contribution from $\hat{D}$ such that it includes (to leading order)
only contributions from generic, isolated and unstable periodic
orbits. This is done since we would like to deduce from the 2D Sinai
billiard on the 2D generic case in which the bouncing--balls are not
present. (In the Sinai billiard, which is concave, there are also
diffraction effects
\cite{PSSU96,PSSU97}, but we did not treat them here.) Thirdly, the
variant of (\ref{eq:dtktau}) for billiards reads:
\begin{equation}
  \left\langle \left| \hat{D}(l) \right|^2 \right\rangle_{l} {\rm d}l =
  \frac{K(\xi)}{4 \pi^2 \xi^2} \ {\rm d}\xi
  \label{eq:dxkxi}
\end{equation}
when $\xi \equiv l/L_{\rm H}$. In figure \ref{fig:dikt.2} we
demonstrate the compliance of the form factor with RMT GOE using the
integrated version of the above relation, and taking into account the
presence of two independent spectra. Fourthly, it is interesting to
know the actual numerical values of $\delta^{\rm (2), 2D}_{\rm
smooth}(k)$ for the $k$ values that we considered. We carried out the
computation, and the results are presented in figure
\ref{fig:ddelta2.2}. It is interesting to observe, that for the entire
range we have $\delta^{\rm (2), 2D}_{\rm smooth}(k) \approx 0.1 \ll
1$, which is very encouraging from an ``engineering'' point of view.

We now turn to the analysis of the 3D Sinai billiard. The billiard is
the free space between a cube of edge $L$ and an inscribed sphere of
radius $R$, where $2R < L$ (see figure \ref{fig:sb2d-sb3d}). We used
$L=1$ and $R=0.2$ and desymmetrized the billiard to the fundamental
element (1/48 of the original one). We calculated the lowest 6697
quantum levels in the interval $0 < k < 281.1$ and the shortest 586965
periodic orbits with length $0 < l < 5$ (the number includes
repetitions and time--reversal conjugates).

To treat the 3D Sinai billiard we need to somewhat modify the
formalism which was presented in the previous section. This is due to
the fact that in 3D the contributions of the various non--generic
bouncing--ball manifolds overwhelm the spectrum
\cite{PS95,Pri97}. Since our goal is to give an indication of the
semiclassical error in generic systems, it is imperative to get rid of
this very strong non--generic effect. The bouncing ball amplitudes are
${\cal O}(k^{(s-1)/2})$ where $s$ is the dimensionality of
bouncing--ball manifold in configuration space. In 3D $s=3$ typically,
which completely overwhelms the contributions from isolated periodic
orbits whose amplitude is ${\cal O}(k^0)$. Even the diffraction
corrections to the bouncing--ball amplitudes in 3D increase as ${\cal
O}(k^{\gamma})$ with $\gamma > 0$. In contrast with the 2D case,
however, it is difficult to subtract the bouncing--ball contributions
analytically for two main reasons. First, there are always infinitely
many bouncing--ball primitive families in the 3D case, while in 2D
there is only a finite number. Indeed, in any finite length interval
there exists only a finite number of bouncing ball lengths, however
their number for billiards in 3D exceeds by far the corresponding
number for billiards in 2D. Second, the semiclassical amplitudes of
the bouncing--balls are proportional to their volume in configuration
space, and it is difficult in general to calculate it
analytically. Thus, to overcome these difficulties we had to device a
special method to cleanse the spectrum from the effect of the
bouncing--balls and from the leading diffractive corrections. This
method relies on the sensitivity of the eigenvalues to the boundary
condition and it is described in detail in \cite{SPSUS95,Pri97}. We
shall briefly describe the essence of the method.

The most general (``mixed") boundary conditions under which the
quantum billiard problem is self--adjoint can be written as
\cite{BB70,SPSUS95}:
\begin{equation}
  \kappa \cos\alpha \, \psi(\vec{r}\,) +
  \sin\alpha \, \partial_{\vec{n}} \psi(\vec{r}\,) = 0 
  \; , \; \; \;
  \vec{r} \in (\mbox{boundary of the billiard})
  \label{eq:mbc}
\end{equation}
where $\vec{n}$ is the normal pointing outside of the billiard, the
angle $\alpha$ interpolates smoothly between Dirichlet ($\alpha = 0$)
and Neumann ($\alpha = \pi / 2$) cases and $\kappa$ has the dimension
of a wavenumber. Note that $\alpha$ and $\kappa$ can be different for
different parts of the boundary. The spectrum depends parametrically
on the boundary parameters, $\{k_n(\alpha, \kappa)\}$, and thus we can
define the counting function $N(k; \alpha, \kappa)$. In
\cite{SPSUS95,Pri97} we discussed the 2D and 3D Sinai billiards,
and showed that if we choose Dirichlet boundary conditions on all the
billiard boundaries {\em excluding} the circle (sphere), and set
$\kappa \ne 0$ on the circle (sphere) then the quantity
\begin{equation}
  \tilde{d}(k; \kappa) \equiv
  \left.
  \frac{\partial N(k; \alpha, \kappa)}
       {\partial \alpha}
  \right|_{\alpha = 0} =
  \sum_n  \left.
   \frac{\partial k_{n}(\alpha, \kappa)}
        {\partial \alpha} \right|_{\alpha = 0}
  \delta ( k - k_n )
  \equiv
  \sum_n v_n \delta ( k - k_n )
  \label{eq:dqm-mbc}
\end{equation}
is to a large extent free of the effects of the bouncing--balls. In
the above $k_n \equiv k_n (\alpha = 0)$ which are the
Dirichlet--everywhere eigenvalues. The corresponding semiclassical
trace formula reads \cite{SPSUS95,Pri97}:
\begin{equation}
  \tilde{d}_{\rm sc}(k; \kappa) =
  ( \mbox{smooth part} ) +
  \sum_{\rm po} A_j B_j \cos (k L_j - \nu_j \pi / 2 )
  \label{eq:derivmbc}
\end{equation}
where
\begin{equation}
  B_j = \frac{2 k}{\kappa L_j}
  \sum_{i=1}^{n_j} \cos\theta_{i}^{(j)}.
\end{equation}
In the above $n_j$ is the number of collisions with the circle
(sphere) of the j'th periodic orbit, and $\{ \theta_{i}^{(j)} \}$ are
the angles of incidence on the boundary with respect to the normal. We
note that (\ref{eq:dqm-mbc}) is a weighted density of states where the
standard $1$ weights of the $\delta$ functions are replaced by $v_n$,
and (\ref{eq:derivmbc}) is a weighted trace formula where the standard
amplitudes $A_j$ are replaced by $A_j B_j$. One can show, that $B_j
\approx v_n$ for long enough (ergodic) orbits.

We shall use $\tilde{d}$ for our purposes as follows. Let us consider
the weighted counting function:
\begin{equation}
  \tilde{N}(k) \equiv
  \int_{0}^{k} {\rm d}k' \  \tilde{d}(k') =
  \sum_{n} v_n \Theta(k-k_n) .
\end{equation}
The function $\tilde{N}$ is a staircase with stairs of variable height
$v_n$. As was explained above, its advantage over $N$ is that it is
semiclassically free of bouncing balls effects (to leading order) and
corresponds only to the generic periodic orbits
\cite{SPSUS95}. Similarly, we construct from $\tilde{d}_{\rm sc}$
the function $\tilde{N}_{\rm sc}$. Having defined $\tilde{N},
\tilde{N}_{\rm sc}$, we proceed in analogy to the Dirichlet case. We form
from $\tilde{N}, \tilde{N}_{\rm sc}$ the functions $\hat{N},
\hat{N}_{\rm sc}$, respectively, by multiplying with a window
function $w(k'-k)$ and then construct the measure $\delta^{\rm (2)}$
as in (\ref{eq:delta2int}). The only difference is that the
normalization of $w$ must be modified to account for the ``velocities"
$v_n$ such as:
\begin{equation}
   \bar{d}^{-1}(k) \sum_n v_n^2 |w(k_n-k)|^2 = 1.
\end{equation}
The above considerations are meaningful provided that the
``velocities'' $v_n$ are narrowly distributed around a well-defined
mean $v(k)$ and we consider a small enough $k$-interval, such that
$v(k)$ does not change appreciably within this interval. Otherwise,
$\delta^{\rm (2)}$ is greatly affected by the fluctuations of $v_n$
(which is undesired) and the meaning of the normalization is
questionable.

\begin{figure}[t]
  \begin{center}
    \vspace*{-0.8cm}
    \leavevmode
    \vspace*{-0.5cm}
    \psfig{figure=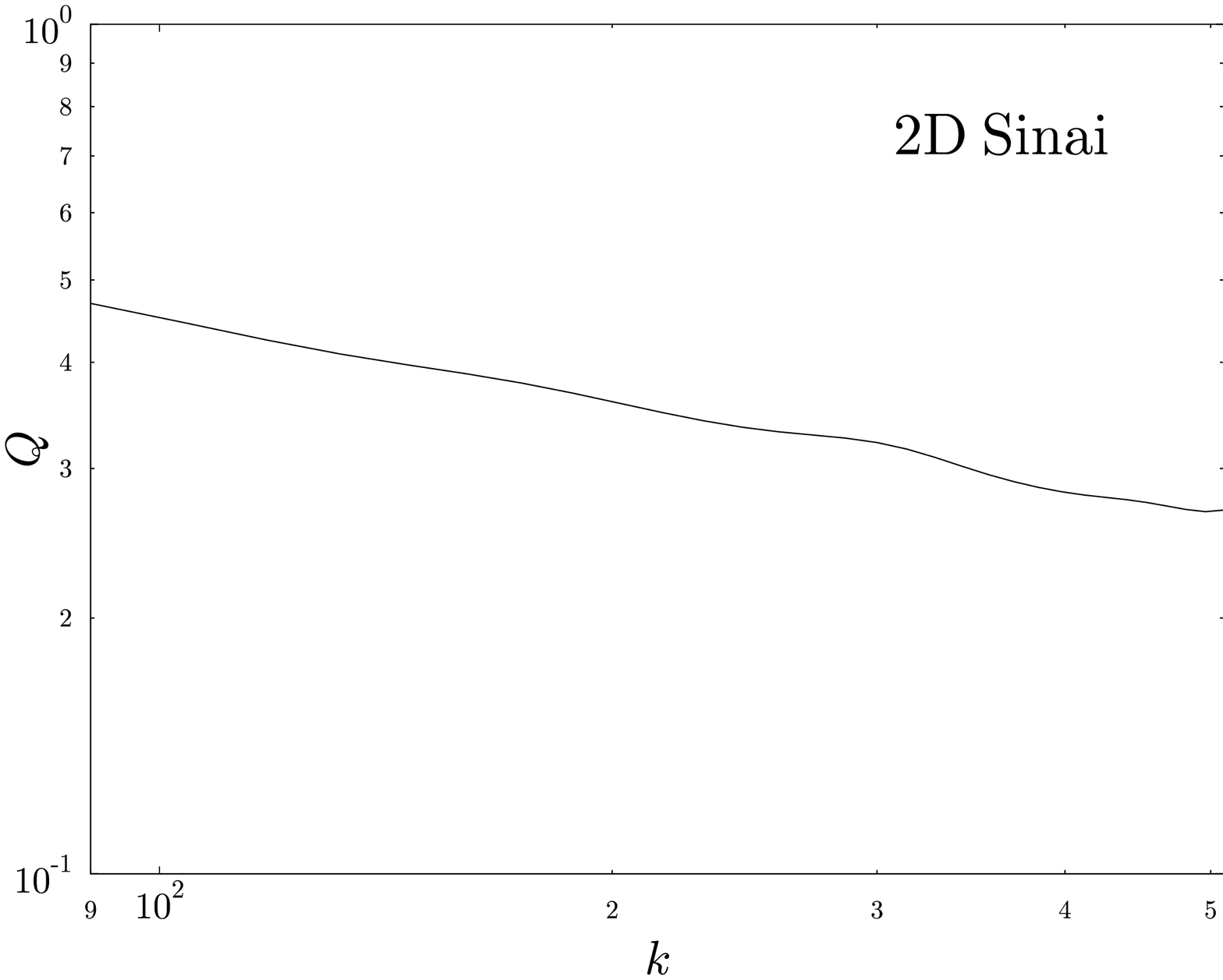,height=9cm}
    \psfig{figure=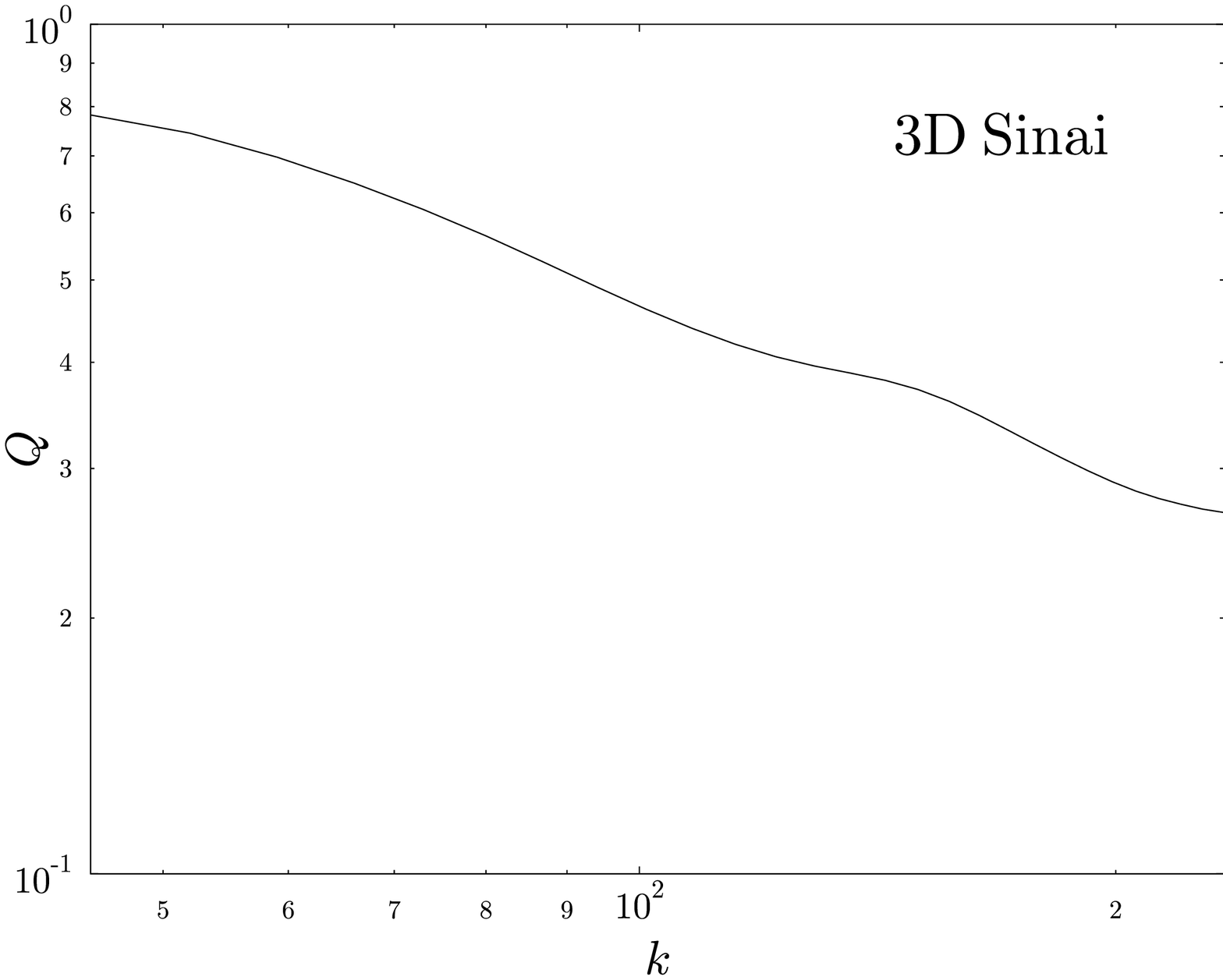,height=9cm}
    \caption{Calculation of $Q \equiv \protect\sqrt{\langle
     v_n^2 \rangle - \langle v_n \rangle^2} / |\langle v_n
      \rangle|$ for quarter 2D Sinai billiard (up)
      and for the desymmetrized 3D Sinai billiard (down).}
    \label{fig:q}
  \end{center}
\end{figure}

\begin{figure}[ht]
  \begin{center}
    \leavevmode
    \psfig{figure=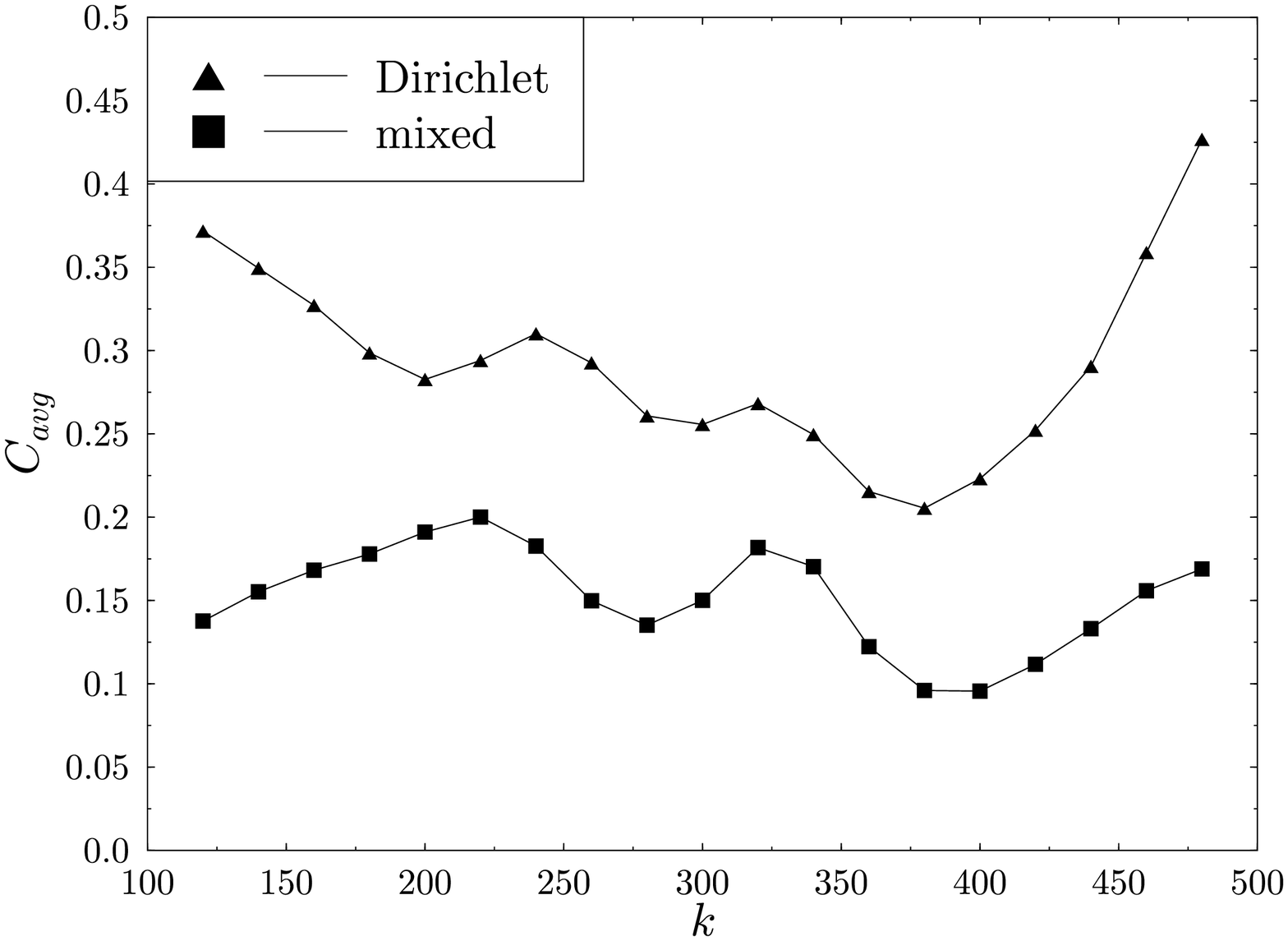,width=12cm}
    \caption{Comparison of $C_{\rm avg}(k)$ for Dirichlet
      and mixed boundary conditions for the quarter 2D
      Sinai billiard. We used a Gaussian
      window with $\sigma = 40$.}
    \label{fig:dmcavg.2}
  \end{center}
\end{figure}

To demonstrate the utility of the above construction using the mixed
boundary conditions, we return to the 2D case. We set $\kappa = 100
\pi$, and note that the spectrum at our disposal for the mixed case 
was confined only to the interval $0 < k < 600$. First of all, we want
to examine the width of the distribution of the $v_n$'s. In figure
\ref{fig:q} we plot the ratio of the standard deviation of $v_n$ to
the mean, averaged over the $k$-axis using a Gaussian window. We use
the same window also in the calculations below. The observation is
that the distribution of $v_n$ is moderately narrow and the width
decreases algebraically as $k$ increases. This justifies the use of
the mixed boundary conditions as was discussed above. One also needs
to check the validity of (\ref{eq:dxkxi}), and indeed we found
compliance with GOE also for the mixed case (results not shown). We
next compare the functions $C(l;k)$ for both the Dirichlet and the
mixed boundary conditions. It turns out, that also in the mixed case
the functions $C(l; k)$ (not shown) fluctuate in $l$ with no special
tendency. The averages $C_{\rm avg}(k)$ for the Dirichlet and mixed
cases are compared in figure \ref{fig:dmcavg.2}. The values in the
mixed case are systematically smaller than in the Dirichlet case which
is explained by the efficient filtering of tangent and close to
tangent orbits that are vulnerable to large diffraction corrections
\cite{PSSU96,PSSU97}. However, from $k = 250$ on the two graphs show
the same trends, and the values of $C_{\rm avg}$ in both cases are of
the same magnitude. Thus, the qualitative behavior of $\delta^{\rm
(2)}_{\rm smooth}$ is shown to be equivalent in the Dirichlet and
mixed cases, which gives us confidence in using $\delta^{\rm (2)}_{\rm
smooth}$ together with the mixed boundary conditions procedure.

\begin{figure}[t]
  \begin{center}
    \leavevmode
    \psfig{figure=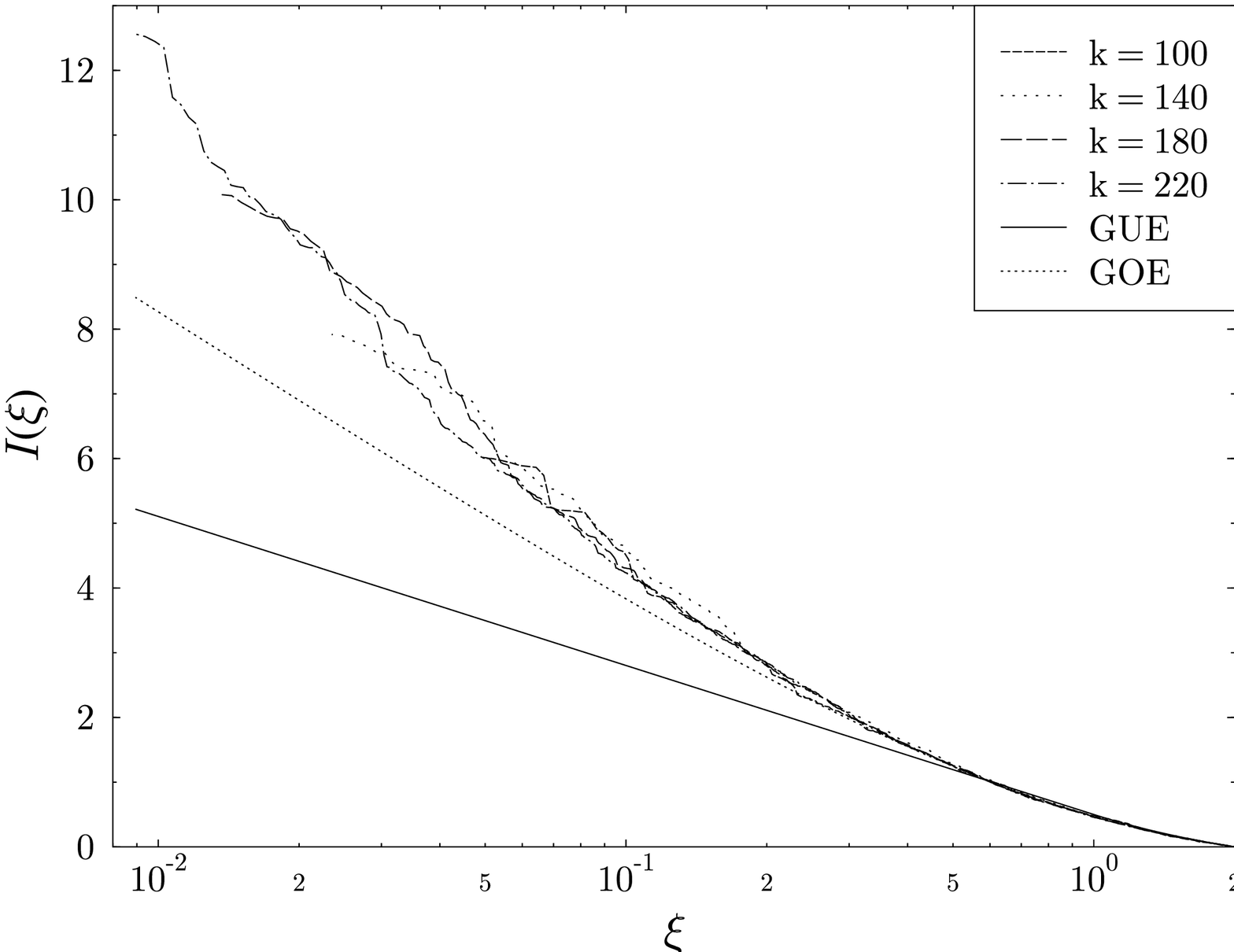,width=12cm}
    \caption{Check of equation (\protect\ref{eq:dxkxi}) for the
      desymmetrized 3D Sinai billiard. The minimal $\xi$ corresponds
      to $L_{\rm erg}=2.5$. The function $I(\xi)$ is defined as in figure
      \protect\ref{fig:dikt.2}. Note the logarithmic scale.}
    \label{fig:mikt.3}
  \end{center}
\end{figure}

\begin{figure}[ht]
  \begin{center}
    \leavevmode
    \psfig{figure=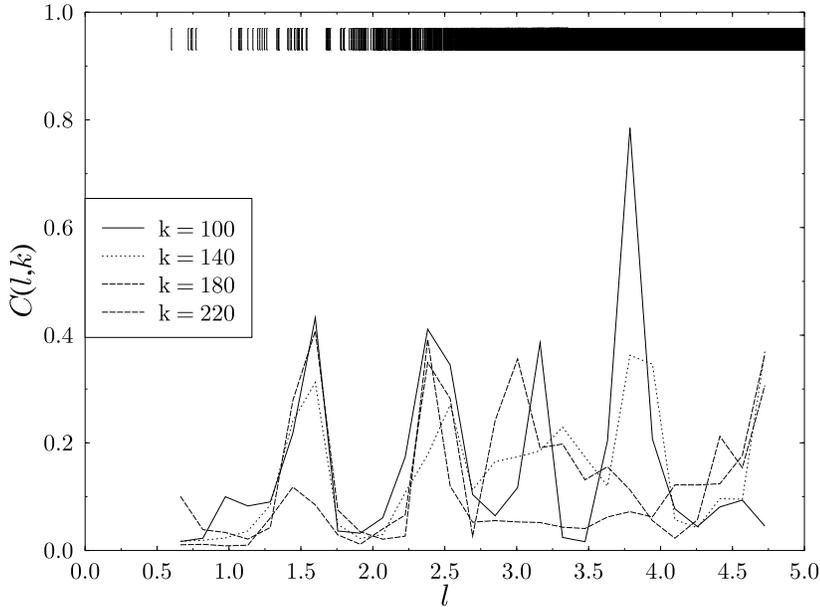,width=12cm}
    \caption{The functions $C(l; k)$ for desymmetrized 3D Sinai
      billiard $L=1, R=0.2$ with mixed boundary conditions. We took a
      Gaussian window with $\sigma=20$, and smoothed over
      $l$-intervals of $\approx 0.3$. The upper vertical bars
      indicate the locations of primitive periodic orbits.}
    \label{fig:mcxavg.3}
  \end{center}
\end{figure}

\begin{figure}[ht]
  \begin{center}
    \leavevmode
    \psfig{figure=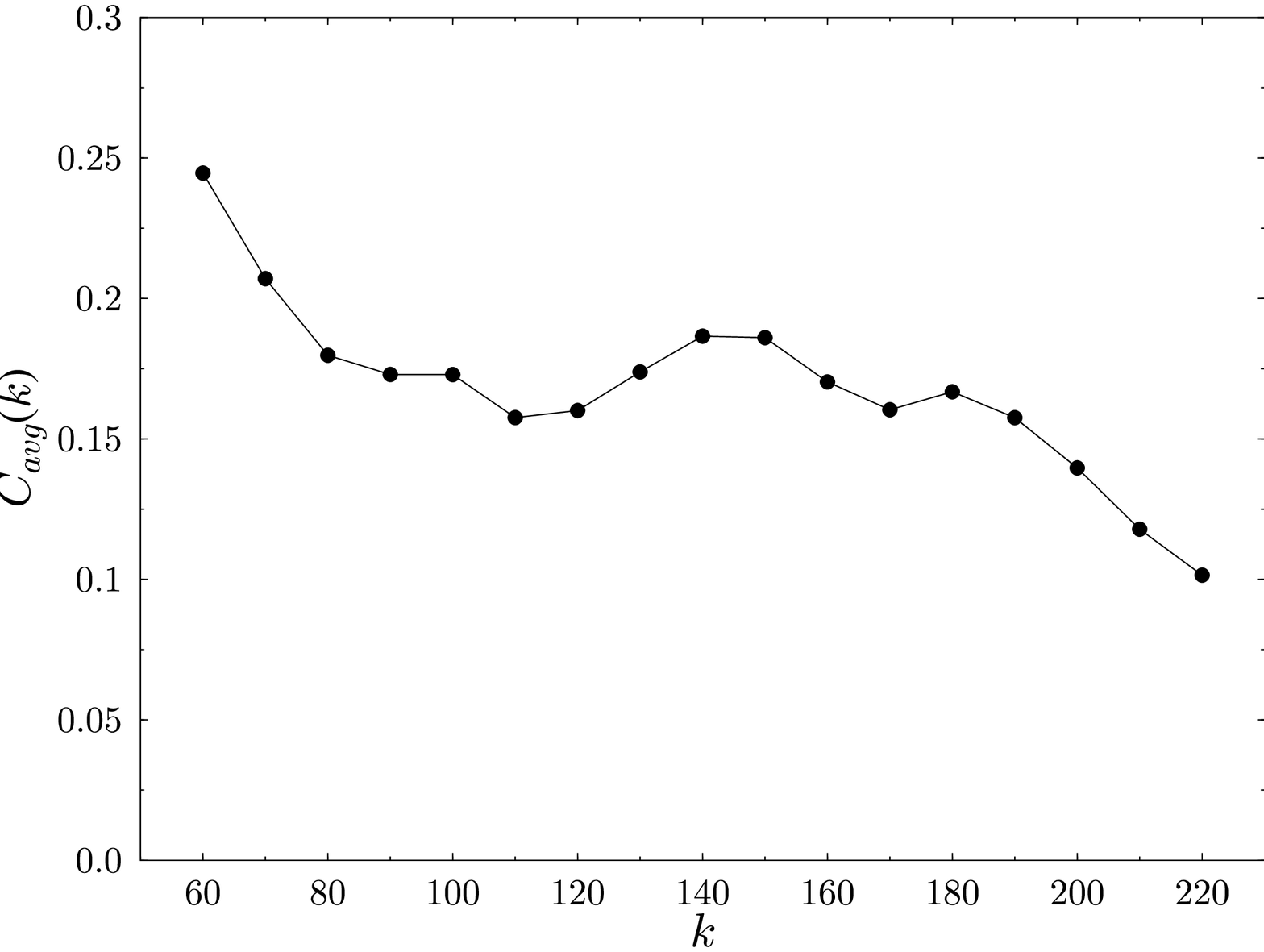,width=12cm}
    \caption{Averaging in $l$ of $C(l; k)$ for 3D Sinai billiard
      as a function of $k$. The averaging was performed in the
      interval $L_{\rm erg}=2.5 < l < 5 = L_{\rm cpu}$.}
    \label{fig:mcavg.3}
  \end{center}
\end{figure}

We finally applied the mixed boundary conditions procedure to compute
$\delta^{\rm (2)}_{\rm smooth}$ for the desymmetrized 3D Sinai with
$L=1, R=0.2$ and set $\kappa = 100$. We first verified that also in
the 3D case the velocities $v_n$ have narrow distribution --- see
figure \ref{fig:q}. Next, we examined equation (\ref{eq:dxkxi}) using
quantal data, and discovered that there are deviations form GOE
(figure \ref{fig:mikt.3}). We have yet no satisfactory explanation of
these deviations, but we suspect that they are caused because the
ergodic limit is not yet reached for the length regime under
consideration due to the effects of the infinite horizon which are
more acute in 3D. Nevertheless, from observing the figure as well as
suggested by semiclassical arguments, it is plausible to assume that
$K(\xi) \propto \xi$ for small $\xi$. Hence, this deviation should not
have any qualitative effect on $\delta^{\rm (2)}$ according to
(\ref{eq:delta2final}). Similarly to the 2D case, the behavior of the
function $C(l; k)$ is fluctuative in $l$, with no special tendency
(figure \ref{fig:mcxavg.3}). If we average $C(l; k)$ over the
universal interval $L_{\rm erg} = 2.5 \leq l \leq L_{\rm cpu} = 5$, we
obtain $C_{\rm avg}(k)$ which is shown in figure
\ref{fig:mcavg.3}. The averages $C_{\rm avg}(k)$ are fluctuating with a
mild decease in $k$, and therefore we can conclude that
\begin{equation}
  {\cal O}(k^{-\beta} \ln k) \leq
  \delta^{\rm (2), 3D}_{\rm smooth}
  \leq {\cal O}(\ln k)
  \label{eq:delta2-3d}
\end{equation}
where the ``optimistic'' measure (leftmost term) corresponds to
$C_{\rm avg}(k) = {\cal O}(k^{-\beta}), \beta > 0$, and the
``pessimistic'' one (rightmost term) is due to $C_{\rm avg}(k) = {\rm
const}$. In other words, the error estimates (\ref{eq:delta2-2d},
\ref{eq:delta2-3d}) for the 2D and the 3D cases, respectively, are the
same, and in sharp contrast to the ``traditional'' error estimate
which predicts that the errors should be different by a factor ${\cal
O}(\hbar^{-1})$. On the basis of our numerical data, and in spite of
the uncertainties which were clearly delineated, we can safely exclude
the ``traditional" error estimate.

\section{Discussion}
\label{sec:discussion}
%
Our main finding was that the upper bound on the semiclassical error
is a logarithmic divergence, which is independent of the dimension
(equations (\ref{eq:delta2-2d}), (\ref{eq:delta2-3d})). In this
respect, there are a few points which deserve discussion.

To begin, we shall try to evaluate $\delta^{\rm (2)}_{\rm smooth}$
using the explicit expressions for the leading corrections to the
semiclassical counting function of 2D generic billiard system, as
derived by Alonso and Gaspard \cite{AG93}:
\begin{equation}
  N(k) = \bar{N}(k) + \sum_{j} \frac{A_j}{L_j} \sin \left[ k L_j +
    \frac{Q_j}{k} + {\cal O}(1/k^2) \right]
\end{equation}
where $A_j$ are the standard semiclassical amplitudes (see
(\ref{eq:gtf})), $L_j$ are the lengths of periodic orbits and $Q_j$
are the $k$-independent amplitudes of the $1/k$ corrections. The
$Q_j$'s are given in \cite{AG93}. We ignored in the above equation the
case of odd Maslov indices. If we calculate from $N(k)$ the
corresponding length spectrum $\hat{D}(l; k)$ using a (normalized)
Gaussian window $w(k'-k) = (1 / \sqrt[4]{\pi \sigma^2}) \exp [
-(k'-k)^2/(2 \sigma^2) ]$, we obtain:
\begin{equation}
  \hat{D}(l; k)
  \approx
  \frac{i \sqrt{\sigma}}{2 \sqrt[4]{\pi}}
  \sum_j \frac{A_j}{L_j}
  \left[
    {\rm e}^{i k (l - L_j) - i \frac{Q_j}{k}}
    {\rm e}^{-(l - L_j)^2 \frac{\sigma^2}{2}} -
    {\rm e}^{i k (l + L_j) + i \frac{Q_j}{k}}
    {\rm e}^{-(l + L_j)^2 \frac{\sigma^2}{2}}
  \right] .
\end{equation}
In the above we regarded the phase ${\rm e}^{i Q_j / k}$ as slowly
varying.  The results of Alonso and Gaspard \cite{AG93} suggest that
the $Q_j$ are approximately proportional to the length of the
corresponding periodic orbits:
\begin{equation}
  Q_j \approx Q L_j .
  \label{eq:mj-lj}
\end{equation}
 We can therefore well--approximate $\hat{D}$ as:
\begin{equation}
  \hat{D}(l; k) \approx 
  \frac{i \sqrt{\sigma}}{2 \sqrt[4]{\pi}} {\rm e}^{-i Q l / k}
  \sum_j \frac{A_j}{L_j} \left[ \cdots \right] =
  {\rm e}^{- i Q l / k} \hat{D}_{\rm sc-GTF} \, ,
\end{equation}
where $\hat{D}_{\rm sc-GTF}$ is the length spectrum which corresponds
to the semiclassical Gutzwiller trace formula for the counting
function (without $1/k$ corrections). We are now in a position to
evaluate the semiclassical error, indeed:
\begin{eqnarray}
  \delta^{\rm (2)}_{\rm smooth}(k)
  & = &
  2 \int_{L_{\rm min}}^{L_{\rm H}} {\rm d}l
  \left| \hat{D}(l; k) - \hat{D}_{\rm sc-GTF}(l; k) \right|^2 =
  \nonumber \\
  & = &
  8 \int_{L_{\rm min}}^{L_{\rm H}} {\rm d}l \,
  \sin^2 \left( \frac{Q l}{2k} \right)
  \left| \hat{D}(l; k) \right|^2 \, .
\end{eqnarray}
If we now use equation (\ref{eq:dxkxi}) and $K(l) \approx g l / L_{\rm
H}$ (which is valid for $l < L_{\rm H}$ for chaotic systems), we get:
\begin{equation}
  \delta^{\rm (2)}_{\rm smooth}(k) 
  \approx 
  \frac{2 g}{\pi^2} \int_{L_{\rm min}}^{L_{\rm H}}
  \frac{{\rm d}l}{l} \sin^2 \left( \frac{Q l}{2 k} \right) =
  \frac{2 g}{\pi^2} \int_{Q L_{\rm min} / (2k)}^{Q L_{\rm H} / (2 k)}
  {\rm d}t \frac{\sin^2 (t)}{t} \, .
  \label{eq:delta2-analytical1}
\end{equation}
For $k \rightarrow \infty$ we have that
\begin{equation}
  \int_{0}^{Q L_{\rm min} / (2k)} {\rm d}t
  \frac{\sin^2 (t)}{t} \approx
  \int_{0}^{Q L_{\rm min} / (2k)}
  {\rm d}t \cdot t = {\cal O}(1/k^2)
\end{equation}
which is negligible, hence we can replace the lower limit in
(\ref{eq:delta2-analytical1}) with 0:
\begin{equation}
  \delta^{\rm (2)}_{\rm smooth}(k) \approx
  \frac{2 g}{\pi^2} \int_{0}^{\frac{Q L_{\rm H}}{2 k}}
  {\rm d}t \frac{\sin^2 (t)}{t} \, .
  \label{eq:delta2-analytical2}
\end{equation}
This is the required expression. The dimensionality enter in
$\delta^{\rm (2)}_{\rm smooth}(k)$ only through the power of $k$ in
$L_{\rm H}$.

Let us apply equation (\ref{eq:delta2-analytical2}) to 2D and 3D
cases. For 2D we have that $L_{\rm H} = A k$ in leading order, where
$A$ is the billiard's area, thus,
\begin{equation}
  \delta^{\rm (2), 2D}_{\rm analytical}(k) \approx
  \frac{2 g}{\pi^2} \int_{0}^{Q A / 2}
  {\rm d}t \frac{\sin^2 (t)}{t} = {\rm const} = {\cal O}(k^0)
  \label{eq:delta2-analytical-2d}
\end{equation}
which means, that the semiclassical error in 2D billiards is of the
order of the mean spacing, and therefore the semiclassical trace
formula is (marginally) accurate and meaningful. This is compatible
with our numerical findings within the limitations of the numerical
fluctuations.

For 3D, the coefficients $Q_j$ were not obtained explicitly, but we
shall assume that they are still proportional to $L_j$
(equation~(\ref{eq:mj-lj})) and thus that
(\ref{eq:delta2-analytical2}) holds.  For 3D billiards $L_{\rm H} = (V
/ \pi) k^2$ to leading order, where $V$ is the billiard's volume. Thus
the upper limit in (\ref{eq:delta2-analytical2}) is $Q V k / (2 \pi)$
which is large in the semiclassical limit. In that case, we can
replace $\sin^2 (t)$ with its mean value $1/2$ and the integrand
becomes essentially $1/t$ which results in:
\begin{equation}
  \delta^{\rm (2), 3D}_{\rm analytical}(k) = {\cal O}(\ln k) \, .
  \label{eq:delta2-analytical-3d}
\end{equation}
That is, in contrast to the 2D case, the semiclassical error diverges
logarithmically and the semiclassical trace formula becomes
meaningless as far as the prediction of individual levels is
concerned. This is compatible with our numerical results within the
numerical dispersion. However, it relies heavily on the assumption
that $Q_j \approx Q L_j$, for which we can offer no justification. We
note in passing, that the logarithmic divergence persists also for $d
> 3$.

Another interesting point relates to integrable systems. It can
happen, that for an integrable system it is either difficult or
impossible to express the Hamiltonian as an explicit function of the
action variables. In that case, we cannot assign to the levels other
quantum numbers than their ordinal number, and the semiclassical error
can be estimated using $\delta^{\rm (2)}$. However, since for
integrable systems $K(\tau) = 1$, we get that:
\begin{equation}
  \delta^{\rm (2), int}_{\rm smooth} 
  \approx \frac{1}{2 \pi^2}
  \int_{\tau_{\rm erg}}^{1} {\rm d}\tau \, 
  \frac{C(\tau)}{\tau^2}.
  \label{eq:delta2-integrable}
\end{equation}
Therefore, for deviations which are comparable to the chaotic cases,
$C(\tau) = {\cal O}(1)$, we get $\delta^{\rm (2), int}_{\rm smooth} =
{\cal O}(\hbar^{1-d})$ which is much larger than for the chaotic case
and diverges for $d \geq 1$.

The formula (\ref{eq:delta2final}) for the semiclassical error
contains semiclassical information in two respects. Obviously,
$C(\tau)$, which is the difference between the quantal and the
semiclassical length spectra contains semiclassical information. But
also, the fact that the lower limit of the integral in
(\ref{eq:delta2final}) is finite is a consequence of semiclassical
analysis. If this lower limit is replaced by $0$, the integral
diverges for finite values of $\hbar$, which is meaningless.
Therefore, the fact that the integral has a lower cutoff, or rather,
that $D$ is exactly $0$ below the shortest period, is a crucial
semiclassical ingredient in our analysis. 

Finally, we consider the case in which the semiclassical error is
estimated with no periodic orbits taken into account. That is, we want
to calculate $\langle | N(E) - \bar{N}(E) |^2 \rangle_{E}$ which is
the number variance $\Sigma^{2}(x)$ for the large argument $x = \Delta
E \, \bar{d}(E) \gg 1$. This implies $C(\tau) = 1$, and using
(\ref{eq:delta2final}) we get that $\delta^{\rm (2)}_{\rm smooth} =
g/(2 \pi^2) \ln (t_{\rm H}/t_{\rm erg})$, which in the semiclassical
limit becomes $g/(2 \pi^2) \ln (t_{\rm H}) = {\cal O}(\ln
\hbar)$. This result is fully consistent and compatible with previous
results for the asymptotic (saturation) value of the number variance
$\Sigma^{2}$ (see for instance \cite{Ber89,BS93,ABS94}). It implies
also that the pessimistic error bound (\ref{eq:delta2-pessimistic})
is of the same magnitude as if periodic orbits were not taken into
account at all. (Periodic orbits improve, however, quantitatively,
since in all cases we obtained $C_{\rm avg} < 1$.) Thus, if we assume
that periodic orbit contributions do not make $N_{sc}$ worse than
$\bar{N}$, then the pessimistic error bound ${\cal O} (\ln \hbar)$ is
the {\em maximal} one in any dimension $d$. This excludes, in
particular, algebraic semiclassical errors, and thus refutes the
traditional estimate ${\cal O} (\hbar^{2-d})$.

\section*{Acknowledgments}
%
This work was finished when HP held a Minerva Post doctoral fellowship
in Freiburg University and US spent a sabbatical leave at the Isaac
Newton Institute, Cambridge. Both authors thank their respective hosts
for the hospitality and support. We would like to acknowledge
discussions with and comments from Michael Berry, Reinhold Bl\"{u}mel,
Doron Cohen, Predrag Cvitanovic, Bruno Eckhardt, Martin Gutzwiller and
Frank Steiner. The work was supported by the Minerva center for
Physics of Complex Systems and by grants from the Israel Science
Foundation.

\section*{References}
%
\bibliographystyle{unsrt}

\end{document}